\documentclass[preprint,12pt]{elsarticle}

\usepackage{lineno}
\usepackage[english]{babel}
\usepackage{fullpage}
\usepackage{amsmath}
\usepackage{amsfonts}
\usepackage{textgreek}
\usepackage{natbib}
\biboptions{compress}
\usepackage{graphicx}
\usepackage{breqn}
\usepackage{undertilde}
\usepackage{appendix}
\usepackage{accents}
\usepackage{enumerate}
\usepackage{caption}
\usepackage{algorithm}
\usepackage{stmaryrd}
\usepackage{color}
\usepackage{wasysym}
\usepackage{mathrsfs}
\usepackage{subcaption}

\usepackage[noend]{algpseudocode}

\usepackage[hidelinks]{hyperref}
\hypersetup{colorlinks,bookmarksopen,linkcolor=blue,citecolor=blue,urlcolor=blue}

\modulolinenumbers[1]

\newcommand*\patchAmsMathEnvironmentForLineno[1]{%
\expandafter\let\csname old#1\expandafter\endcsname\csname #1\endcsname
\expandafter\let\csname oldend#1\expandafter\endcsname\csname end#1\endcsname
\renewenvironment{#1}%
{\linenomath\csname old#1\endcsname}%
{\csname oldend#1\endcsname\endlinenomath}}%
\newcommand*\patchBothAmsMathEnvironmentsForLineno[1]{%
\patchAmsMathEnvironmentForLineno{#1}%
\patchAmsMathEnvironmentForLineno{#1*}}%
\AtBeginDocument{%
\patchBothAmsMathEnvironmentsForLineno{equation}%
\patchBothAmsMathEnvironmentsForLineno{align}%
\patchBothAmsMathEnvironmentsForLineno{flalign}%
\patchBothAmsMathEnvironmentsForLineno{alignat}%
\patchBothAmsMathEnvironmentsForLineno{gather}%
\patchBothAmsMathEnvironmentsForLineno{multline}%
}
\definecolor{myred}{RGB}{222,45,38}
\definecolor{myblue}{RGB}{0,115,189}
\definecolor{mygreen}{RGB}{49,156,54}
\newcommand{\bs}[1]{{\boldsymbol{#1}}}

\graphicspath{{figs/}}

\journal{Mech. Mater.}

\newcommand{\sgn}[1]{\mathrm{sgn}\left(#1\right)}

\biboptions{authoryear}

\begin{document}

\begin{frontmatter}

\title{The Peierls--Nabarro finite element model in two-phase microstructures -- a comparison with atomistics\tnoteref{titlefoot}}
\tnotetext[titlefoot]{The post-print version of this article is published in \emph{Mech. Mater.}, \href{https://doi.org/10.1016/j.mechmat.2020.103555}{10.1016/j.mechmat.2020.103555}}

\author[TUe]{F.~Bormann}
\ead{franz.bormann@gmail.com}
\author[CTU]{K.~Mike\v{s}}
\ead{Mikes.Karel.1@fsv.cvut.cz}
\author[TUe]{O.~Roko\v{s}\corref{correspondingauthor}}
\ead{O.Rokos@tue.nl}
\author[TUe]{R.H.J.~Peerlings}
\ead{R.H.J.Peerlings@tue.nl}

\address[TUe]{Department of Mechanical Engineering, Eindhoven University of Technology, PO Box 513, 5600 MB Eindhoven, The Netherlands}
\address[CTU]{Department of Mechanics, Faculty of Civil Engineering, Czech Technical University in Prague,
Th\'{a}kurova~7, 166~29 Prague~6, Czech Republic}

\cortext[correspondingauthor]{Corresponding author.}

%
\begin{abstract}
	This paper evaluates qualitatively as well as quantitatively the accuracy of a recently proposed Peierls--Nabarro Finite Element (PN-FE) model for dislocations by a direct comparison with an equivalent molecular statics simulation. 
	To this end, a two-dimensional microstructural specimen subjected to simple shear is considered, consisting of a central soft phase flanked by two hard-phase regions. A hexagonal atomic structure with equal lattice spacing is adopted, the interactions of which are described by the Lennard--Jones potential with phase specific depths of its energy well. During loading, edge dislocation dipoles centred in the soft phase are introduced, which progress towards the phase boundaries, where they pile up. Under a sufficiently high external shear load, the leading dislocation is eventually transmitted into the harder phase. 
	The homogenized PN-FE model is calibrated to an atomistic model in terms of effective elasticity constants and glide plane properties as obtained from simple uniform deformations. To study the influence of different formulations of the glide plane potential, multiple approaches are employed, ranging from a simple sinusoidal function of the tangential disregistry to a complex model that couples the influence of the tangential and the normal disregistries. 
	The obtained results show that, qualitatively, the dislocation structure, displacement, strain fields, and the dislocation evolution are captured adequately. The simplifications of the PN-FE model lead, however, to some discrepancies within the dislocation core. Such discrepancies play a dominant role in the dislocation transmission process, which thus cannot quantitatively be captured properly. 
	Despite its simplicity, the PN-FE model proves to be an elegant tool for a qualitative study of edge dislocation behaviour in two-phase microstructures, including dislocation transmission, although it may not be quantitatively predictive.
\end{abstract}

\end{frontmatter}

%
%
\section{Introduction}
%
Over the past decades, the Peierls--Nabarro (PN) model~\citep{Hirth1982} has gained popularity in the dislocation community due to its ability to model dislocations at the atomistic scale while using a continuum framework. In its classical form, an infinite and homogeneous crystal is split into two linear-elastic regions connected by a glide plane. Along this glide plane, a relative tangential displacement (or tangential disregistry) is allowed, which is mapped onto an intrinsic misfit energy. In the classical PN model, the adopted glide plane potential is a periodic function which is based on the Frenkel sinusoidal function~\citep{Hirth1982}. Minimising the total free energy, consisting of the elastic strain energy of the bulk and the misfit energy of the glide plane, leads to an arctan-type disregistry profile with satisfactory properties in terms of non-singular stress field, total energy, and Peierls energy as well as Peierls stress. \par
%
The PN model furthermore excels through its versatility when solved numerically. In this fashion, the limitations of the classical approach can be overcome. For instance, the influence of heterogeneous crystals on dislocation obstruction at phase boundaries, as well as on dislocation transmission, can readily be modelled, see, e.g., \citep{Anderson2001, Bormann2018}.
%
It has, however, been pointed out by some authors that, with the utilisation of the simple sine function, the PN model is limited in its quantitative description of dislocations in real crystals \citep{Schoeck2005}. A variety of extensions has therefore been suggested in the literature, as follows.\par
%
A significant increase in the accuracy of the misfit energy compared to the simple one-dimensional (1D) sinusoidal function can be achieved by the Generalized Stacking Fault Energy (GSFE) surface, introduced initially by~\cite{Vitek1968}. It provides a full 2D-energy landscape (in contrast to the 1D sinusoidal function), intrinsic to the crystal considered. With this extension, the PN model is capable of describing mixed dislocations, splitting of dislocations into Shockley partials, and the related recombination energy of partials, see, e.g., \citep{Schoeck2005, Schoeck2001, Wang2011}. The GSFE surface is commonly obtained from Molecular Statics~(MS) calculations or directly from Density Functional Theory (DFT) \citep{Kaxiras1993, Su:2019}. It has been shown recently that for a bilayer system, the solutions of the PN model with the GSFE extension are asymptotically close to the full atomistic model~\citep{Luo2017}.
Yet, the GSFE is limited to tangential (i.e., in-plane) disregistry only. Lifting this constraint and introducing an additional normal (i.e., out-of-plane) disregistry-dependency improves the dislocation description even further~\citep{Sun1993, Bulatov1997, Xiang2008}.
As the misfit energy resides between the atoms located above and below the glide plane (recall that the glide plane often reduces to a zero-thickness interface), it has been suggested to subtract the linear elastic part of the misfit energy~\citep{Sun1993}. This correction leads to a larger activation energy \cite[as discussed by][]{Xu2000a}, and to a decrease of the Peierls stress, cf.~\cite{Xu2003}. 
The inclusion of elastic anisotropy is an important feature for dislocations in anisotropic crystals \citep{Xiang2008, Eshelby1949}. Further PN model extensions include non-local formulations reflecting the discreteness of the underlying atomic lattice~\citep{Bulatov1997, Liu2017}, or an additional gradient energy term~\citep{Wang2015b} motivated by DFT calculations~\citep{Iyer2015}.\par
%
%
Various authors have studied the accuracy and predictive power of the PN model in comparison with atomistic simulations. \cite{Sydow1999} analysed Shockley partial dislocations in Pd and found a relatively good agreement of the PN model. More recently, \cite{Dai2014} unveiled a high accuracy of the generalized PN model in capturing the structure and energy of low-angle grain boundaries and near-twin grain boundaries for (111) twist boundaries in Al, Cu and Ni. \cite{Mianroodi2016} compared the generalised PN model of \cite{Xiang2008}, the phase-field dislocation dynamics model of \cite{Hunter2011}, the phase-field based models of \cite{Shen2004}, and \cite{Mianroodi2015} with MS simulations, and found a rather good agreement in terms of the dissociation of a dislocation dipole in fcc single crystals in Al and Au. Considering a non-local formulation of the PN model, \cite{Liu2017} presented a good predictive capability in terms of the dislocation core structure and of the Peierls stress in Fe and Cu, whereas~\cite{Xu:2019b} compared different continuum approaches for modelling of mixed-type dislocations in Al. \cite{Xu:2019c} furthermore compared various Fourier series approximations of the GSFE in Au and Al. Interactions of dislocations with interfaces in two-phase microstructures have recently been studied, e.g., by~\cite{Mianroodi:2019}.\par
This paper aims at providing the qualitative and quantitative comparison of the predictions made by the recently proposed Peierls--Nabarro finite element (PN-FE) model and by MS simulations. The focus of this comparison is on the pile-up evolution of edge dislocations in a two-phase microstructure under an increasing external shear load and the eventual dislocation transmission across the phase boundary. This manuscript builds on our previous work~\citep{Bormann2018}, by extending and validating the therein proposed methodology against a fully discrete MS model. The model problem considered here is limited to a 2D lattice to reduce the complexity of the 3D reality and to facilitate a sharp comparison. Different complexities of the glide plane potential are considered, ranging from a simple sinusoidal function up to a fully coupled function of the tangential disregistry~$\Delta_t$ as well as the normal disregistry~$\Delta_n$. Their performance, and predictive capabilities, will be assessed critically.\par
%
%
The problem considered employs a 2D microstructure consisting of two phases. A soft central phase (Phase~A) is flanked by two hard-phase regions (Phase~B). Both phases have a hexagonal atomic lattice structure with homogeneous spacing. The specimen is subjected to a shear deformation applied through the external boundary, which induces in a defect-free configuration a state of uniform shear stress. Edge dislocation dipoles are generated in the soft Phase~A on a glide plane perpendicular to the coherent and non-damaging phase boundaries. Due to the phase contrast, the dislocations pile up at the phase boundary where eventually, under a sufficiently high external shear load, the leading dislocation is transmitted into the harder Phase~B.\par 
%
%
The paper is divided into four sections as follows. In Section~\ref{section:Problem_Statement}, the considered problem is described in detail and the basics of the MS and PN-FE models are briefly recalled. The different glide plane potentials employed for the purpose of comparison are subsequently introduced in Section~\ref{sect:GP-models}. Individual solutions of the PN-FE model for a single dislocation dipole and a dipole pile-up are compared with the results of the MS model in Section~\ref{section:Comparison}. The paper closes with a summary and discussion in Section~\ref{section:Discussion}.

%
%
%
%
\section{Problem statement}
\label{section:Problem_Statement}
The geometry of the benchmark problem employed throughout this paper is first specified in  Section~\ref{sect:geometry}, whereas the basics of the atomistic MS calculations along with the prescribed boundary conditions and used potentials are detailed in Section~\ref{sect:MS}. The continuum PN-FE framework itself is described in Section~\ref{sect:pnfe}, along with the required definitions and numerical values of the effective quantities in terms of constitutive linear-elastic constants and glide plane potential.
%
%
\subsection{Atomistic system}
\label{sect:geometry}
\begin{figure}[htbp]
	\centering
	\includegraphics[scale=1]{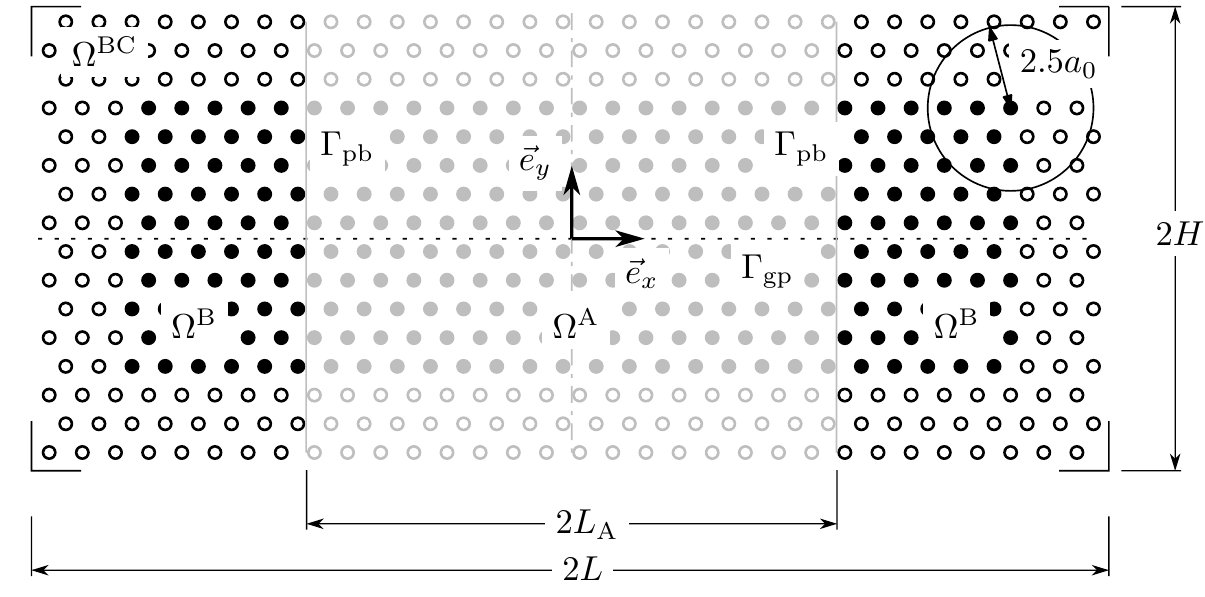}	
	\caption{A sketch of the atomistic representation of a two-phase microstructure used for the simulations of edge dislocation dipoles interacting with a coherent phase boundary. The entire domain ($\Omega$, all atoms) consists of two phases, Phase~A ($\Omega^{\mathrm{A}}$, grey atoms) and Phase~B ($\Omega^{\mathrm{B}}$, black atoms), and of the boundary domain ($\Omega^{\mathrm{BC}}$, white atoms). The positions of all atoms situated inside~$\Omega^{\mathrm{BC}}$ are prescribed to induce a shear deformation.}
	\label{fig:Model}
\end{figure}

A two-phase microstructure, as illustrated in Figure \ref{fig:Model} by a downsized example, is considered. It consists of a soft Phase~A that is flanked by a harder Phase~B. Both phases have an identically oriented hexagonal lattice structure of equal spacing. The crystal orientation is chosen such that one set of glide planes~$\Gamma_{\mathrm{gp}}$ is oriented perpendicular to the coherent phase boundaries~$\Gamma_{\mathrm{pb}}$. Through the external boundary, a shear deformation is induced, which would correspond in an ideal, defect-free, and linear-elastic specimen to a state of homogeneous shear stress~$\tau$. Stable dislocation dipoles are initialised in the centre of the Phase~A, which under increasing shear load move towards the phase boundary. Due to the phase contrast, the dislocations get obstructed at the phase boundaries and gradually pile up. Eventually, under a sufficiently high external shear load, the leading dislocations are transmitted into the harder Phase~B. 
%
%
\subsection{Molecular statics problem}
\label{sect:MS}
Full discrete MS simulations are performed, which serve two purposes. First, they provide the reference solution of the above-specified problem to which the PN-FE model is compared in Section~\ref{section:Comparison}. Second, atomistic simulations of simple, uniform, single-phase problems provide the homogenised material properties of the corresponding continuum PN-FE model, such as linear elasticity constants and glide plane properties.

For studying edge dislocation versus phase boundary interaction, the problem domain~$\Omega = \Omega^{\mathrm{A}} \cup \Omega^{\mathrm{B}}$ is considered, as sketched in Figure~\ref{fig:Model}, with the regions occupied by the two Phases $\mathrm{A}$ and~$\mathrm{B}$ specified as
\begin{align}
\Omega^{\mathrm{A}} &= \left\{ \vec{x} \in \mathbb{R}^2: | x | \leq L_{\mathrm{A}}, | y | \leq H \right\} \\
\Omega^{\mathrm{B}} &= \left\{ \vec{x} \in \mathbb{R}^2: L_{\mathrm{A}} \leq | x | \leq L , | y | \leq H \right\}
\end{align}
where~$\vec{x} = x\vec{e}_x + y\vec{e}_y$ is the position vector. The hexagonal lattice with spacing~$a_0$ is initiated such that a stress-free reference state results (i.e., $\bs{\sigma} = \bs{0}$, cf. Eq.~\eqref{eq:Virialstress} below). Each atom~$\alpha$ located inside the domain~$\Omega$ is stored in an index set~$N = N^{\mathrm{A}} \cup N^{\mathrm{B}}$, where
\begin{align}
N^{\mathrm{A}} &= \{ \alpha \in N : \vec{r}_0^{\,\alpha} \in \Omega^{\mathrm{A}} \} \\
N^{\mathrm{B}} &= \{ \alpha \in N : \vec{r}_0^{\,\alpha} \in \Omega^{\mathrm{B}} \}
\end{align}
and~$\vec{r}_0^{\,\alpha} = r_{0x}^{\alpha}\vec{e}_x + r_{0y}^{\alpha}\vec{e}_y$ denotes the spatial position of an atom~$\alpha$ in the reference configuration.\par
\begin{table}
	\caption{Parameters of the pair potentials corresponding to the two Phases~A and~B, as a function of the material contrast ratio~$\rho = 1.4$.}
	\centering
	\begin{tabular}{c|c|c|c|c}
		Parameter & $\varepsilon/\varepsilon^{\mathrm{A}}$ & $r_{\mathrm{m}}/r_{\mathrm{m}}^{\mathrm{A}}$ & $a_0/r_{\mathrm{m}}^{\mathrm{A}}$ & $r_{\mathrm{cut}}/a_0$ \\\hline
		Phase~$\mathrm{A}$ & $1$ & $1$ & $0.99296702$ & $2.5$ \\
		Phase~$\mathrm{B}$ & $\rho$ & $1$ & $0.99296702$ & $2.5$
	\end{tabular}
	\label{tab:ljsf}
\end{table}
The mechanical behaviour of the system is governed by its total potential energy~$\mathcal{V}$, defined as the sum of all pair potentials~$\phi^{\alpha\beta}(r^{\alpha\beta})$, where~$r^{\alpha\beta} = \| \vec{r}^{\,\alpha\beta} \|_2$ denotes the Euclidean distance between a pair of atoms~$\alpha$ and~$\beta$. A shifted Lennard--Jones potential is employed, with the cut-off radius~$r_{\mathrm{cut}}$. Standard notation is adopted for the unshifted Lennard--Jones potential, i.e., the well-depth is~$\varepsilon$ whereas the distance to the potential minimum is~$r_{\mathrm{m}}$; for further details see, e.g., \citep{Tadmor2011}. Introducing a material contrast ratio $\rho$, the pair potential of Phase~B is set to~$\phi^{\mathrm{BB}} = \rho\phi^{\mathrm{AA}}$. The constitutive parameters associated with the individual phases are summarised in Table~\ref{tab:ljsf}. The interaction potential across the phase boundary is considered as the average of that of the two individual phases, i.e., $\phi^{\mathrm{AB}}=\frac{1}{2}(\phi^{\mathrm{AA}}+\phi^{\mathrm{BB}})$. The material properties are not calibrated to any particular material, and the lattice considered is a model hexagonal crystal. All parameters of the employed LJ potential are thus normalized according to Tab.~\ref{tab:ljsf} with respect to~$r_{\mathrm{m}}^{\mathrm{A}}$ and with respect to~$\varepsilon^{\mathrm{A}}$.\par
The boundary conditions applied on~$\Omega^{\mathrm{BC}}$ are chosen such that in an ideal, defect-free, and linear-elastic specimen, a state of constant shear stress~$\tau$ would result. In the discrete MS model te stress is not perfectly uniform due to the heterogeneities and non-linearities, although in the early stages of loading the stress state is fairly close to a constant. Assuming a phase-wise homogeneous and linear-elastic response, the corresponding shear strains read~$t\overline{\tau}/\mu^{\mathrm{A}}$ in Phase~A and~$t\overline{\tau}/\rho\mu^{\mathrm{A}}$ in Phase~B. Here, $\overline{\tau}$ is the target shear load, $\mu^{\mathrm{A}}$ the homogenised shear modulus of the Phase~A (cf. Section \ref{sect:pnfe} below), and~$t \in [0,1]$ a pseudo-time parametrising the evolution of the otherwise rate-independent system (i.e., zero temperature is assumed). In order to introduce the shear deformation, a layer of atoms~$N^{\mathrm{BC}} \subset N$ is displaced accordingly. The thickness of the layer~$N^{\mathrm{BC}}$ is chosen in accordance with the cut-off radius~$r_{\mathrm{cut}}$ to eliminate any surface effects. Atom positions are prescribed as
\begin{align}
\label{eq:bcs-a}
\vec{r}^{\,\alpha} &= \vec{r}_0^{\,\alpha} + \left[t\frac{\overline{\tau}}{\mu^{\mathrm{A}}}\,r_{0x}^\alpha\right]\,\vec{e}_y, \quad \text{for} \quad \alpha \in N^{\mathrm{A}} \cap N^{\mathrm{BC}}\\
\label{eq:bcs-b}
\vec{r}^{\,\alpha} &= \vec{r}_0^{\,\alpha} + \frac{t\overline{\tau}}{\rho\mu^{\mathrm{A}}}\left[r_{0x}^\alpha-\left(1-\rho\right)L_{\mathrm{A}}\,\sgn{r_{0x}^\alpha}\right]\,\vec{e}_y, \quad \text{for} \quad \alpha \in N^{\mathrm{B}} \cap N^{\mathrm{BC}}
\end{align}
In order to predict the mechanical behaviour of the specimen, the total potential energy is minimised at each time step~$t_k$,
\begin{equation}
\utilde{r}(t_k) \in \underset{{\scriptstyle\utilde{q}}\in Q_k}{\text{arg min}}\ \mathcal{V}(\utilde{q})
\label{eq:ms_min}
\end{equation}
The configuration space at a time step~$t_k$ is denoted as~$Q_k \subseteq \mathbb{R}^{2n_{\mathrm{ato}}}$, and reflects any prescribed atom displacements (recall Eqs.~\eqref{eq:bcs-a} and~\eqref{eq:bcs-b}). The minimisation problem is solved using the Trust-region methodology \cite[cf., e.g.,][]{Conn2000a}, which has been implemented within an in-house code. 

Because multiple glide planes may be activated due to random perturbations and round-off errors, the numerical solver is initialised towards the preferred glide plane~$\Gamma_{\mathrm{gp}} = \{ \vec{x} \in \mathbb{R}^2 : y = 0 \}$. In particular, the resolved shear stress at the predefined position of the source, $\tau_{\mathrm{res}}$, is monitored, and once a critical threshold value~$\tau_{\mathrm{nuc}}$ is exceeded, i.e.~when $\tau_{\mathrm{res}} > \tau_{\mathrm{nuc}}$, a new dislocation dipole centred on the glide plane~$\Gamma_{\mathrm{gp}}$ is initialized through a rescaled Volterra displacement field \cite[see, e.g.,][]{Tadmor2011}
\begin{equation}
\begin{aligned}
\vec{u}^{\mathrm{V}}(\vec{x}) &= \frac{C_{\mathrm{m}}}{2\pi} \left[-\tan^{-1}\frac{x}{y} + \frac{xy}{2(1 - \nu)(x^2 + y^2)}\right]\vec{e}_x \\
&\, -\frac{C_{\mathrm{m}}}{2\pi} \left[\frac{1 - 2\nu}{4(1 - \nu)}\ln(x^2 + y^2) + \frac{x^2 - y^2}{4(1 - \nu)(x^2 + y^2)}\right]\vec{e}_y,
\end{aligned}
\quad \vec{x} \in \mathbb{R}^2
\label{eq:volterra}
\end{equation}
where~$C_{\mathrm{m}}$ is the magnitude of the perturbation (chosen as~$C_{\mathrm{m}} = a_0/2$), whereas~$\nu$ is Poisson's ratio (obtained as a part of the homogenized lattice properties described below in Section~\ref{sect:pnfe}). The displacement field of Eq.~\eqref{eq:volterra} is used to perturb the current relaxed configuration at a time step~$t_k$, i.e.,
\begin{equation}
\vec{r}^{\,\mathrm{init},\alpha}(t_{k+1}) = \vec{r}^{\,\alpha}(t_k) + \vec{u}^{\mathrm{V}}(\vec{r}_0^{\,\alpha} - \ell\vec{e}_x) - \vec{u}^{\mathrm{V}}(\vec{r}_0^{\,\alpha} + \ell\vec{e}_x), \quad \text{for} \quad \alpha \in N \backslash N^{\mathrm{BC}}
\label{eq:init}
\end{equation}
which is used as an initial guess for the minimization algorithm in the consecutive time step~$t_{k+1}$. Note that at the interface between~$\Omega^{\mathrm{A}} \cup \Omega^{\mathrm{B}}$ and~$\Omega^{\mathrm{BC}}$, no artificial jump is created, since the perturbation Volterra displacement field decays rapidly with distance from the dislocation core. To avoid unnecessary computing expenses associated with propagating a dislocation from its initial to its equilibrium position, the dislocation's equilibrium position~$\ell$ is estimated. To this end, the resolved Virial shear stress~$\tau(x)$ superposed with the analytical shear stresses~$\tau_{\mathrm{n}}^\pm(x)$ of the to-be-nucleated dislocation pair according to~\cite{Head1953-Edge} is used. Equilibrium of the combined shear stresses allows one to determine~$\ell$, by solving it from
\begin{equation}
\tau(\ell) + \tau^+_{\mathrm{n}}(\ell) + \tau^-_{\mathrm{n}}(\ell) = 0
\label{eq:nucleation}
\end{equation}
At the same time, it is required that~$\ell \in [0,\ell_{\mathrm{max}}]$, where
\begin{equation}
\ell_{\mathrm{max}} < \left\{
\begin{array}{@{}l@{}l}
L_{\mathrm{A}}, & \quad \mbox{if} \quad j = 0 \\
\min(x_j), & \quad \mbox{if} \quad j > 0
\end{array}
\right.
\end{equation}
and where~$x_j$, $j = 1, \dots, n$, are the distances of the~$n$ already existing dislocations on the glide plane~$\Gamma_{\mathrm{gp}}$ measured from the origin. The estimated dislocation position~$\ell$ is subsequently used in Eq.~\eqref{eq:init} to initialize the dislocation dipole in the next time step. To ensure that early dislocations are captured properly, the critical nucleation stress~$\tau_{\mathrm{nuc}}$ is chosen small in the MS simulations, unlike the PN-FE model (cf. Section~\ref{sect:pnfe} below).

For later reference and for comparison with the PN-FE model, the definition of the pointwise Virial stress~$\bs{\sigma}$ is recalled
\begin{equation}
\sigma_{ij} = \frac{1}{2V} \sum_{\substack{\alpha,\beta\\\alpha\neq\beta}}\phi'(r^{\alpha\beta})\frac{r_i^{\alpha\beta}r_j^{\alpha\beta}}{r^{\alpha\beta}}
\label{eq:Virialstress}
\end{equation}
where~$V$ denotes the area of the Voronoi cell among the nearest neighbours multiplied by a virtual thickness~$r_{\mathrm{m}}^{\mathrm{A}}$, and~$r_i^{\alpha\beta}$ the components of the~$\vec{r}^{\,\alpha\beta}$ vectors.
%
%
\subsection{Peierls--Nabarro finite element (PN-FE) model}
\label{sect:pnfe}
\begin{figure}
	\centering
	\includegraphics[scale=1]{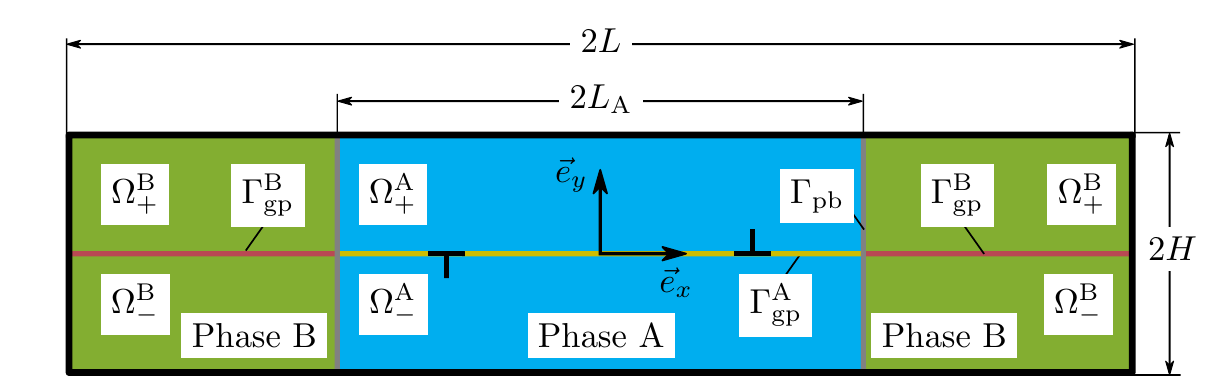}	
	\caption{Continuum model for edge dislocation dipoles interacting with the phase boundary in a two-phase microstructure. Note that although the lines representing~$\Gamma_{\mathrm{pb}}$ and~$\Gamma_{\mathrm{gp}}$ have a certain thickness in the sketch, computationally they are considered as having zero in-plane thickness.}
	\label{fig:FEM-Model}
\end{figure}

The atomistic problem as specified in Figure~\ref{fig:Model} is translated to the PN-FE framework, introduced by \cite{Bormann2018}, in a straightforward manner. Let~$\Omega$ be the two-phase microstructure under consideration, cf. Figure~\ref{fig:FEM-Model}. Similarly to the atomistic model, it consists of two regions~$\Omega^{\mathrm{A}}$ and~$\Omega^{\mathrm{B}}$, that are separated by perfectly bonded and coherent phase boundaries~$\Gamma_{\mathrm{pb}}$ perpendicular to~$\vec{e}_x$, i.e.,
\begin{align}
\Omega^{\mathrm{A}} &= \left\{ \vec{x} \in \mathbb{R}^2: | x | \leq L_{\mathrm{A}}, | y | \leq H \right\} \\
\Omega^{\mathrm{B}} &= \left\{ \vec{x} \in \mathbb{R}^2: L_{\mathrm{A}} \leq | x | \leq L, | y | \leq H \right\} \\
\Omega &= \Omega^{\mathrm{A}} \cup \Omega^{\mathrm{B}} \\
\partial\Omega &= (\partial\Omega^{\mathrm{A}}\backslash\Gamma_{\mathrm{pb}})\cup(\partial\Omega^{\mathrm{B}}\backslash\Gamma_{\mathrm{pb}})
\end{align}	
In~$\Omega$ lies a single glide plane~$\Gamma^i_{\mathrm{gp}}$, perpendicular to~$\vec{e}_y$, that splits each phase~$\Omega^i$, $i\in\left\{\mathrm{A,B}\right\}$, into two subdomains~$\Omega^i_\pm$ as follows:
\begin{align}
\Omega^i &= \Omega^i_+ \cup \Omega^i_- \\
\partial\Omega^i &= (\partial\Omega^i_+\backslash\Gamma^i_{\mathrm{gp}})\cup(\partial\Omega^i_-\backslash\Gamma^i_{\mathrm{gp}})
\end{align}	

\begin{figure}
	\centering
	\begin{subfigure}{0.49\textwidth}
		\centering
		\caption{}
		\includegraphics[scale=1]{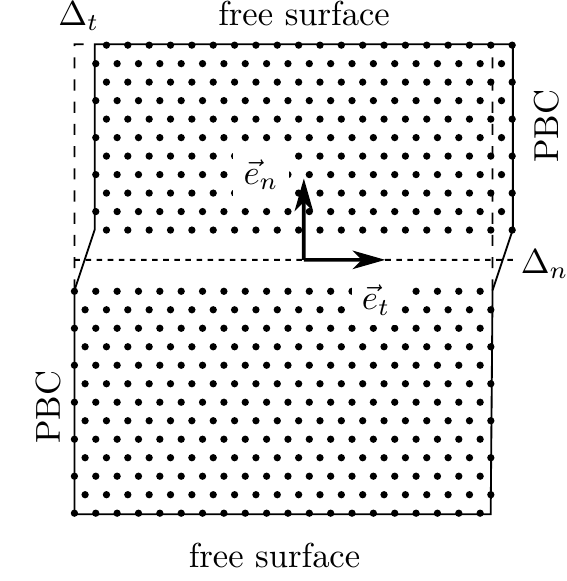}
		\label{fig:msbcs_b}
	\end{subfigure}
	\begin{subfigure}{0.49\textwidth}
		\centering
		\caption{}
		\vspace{14pt}
		\includegraphics[scale=1]{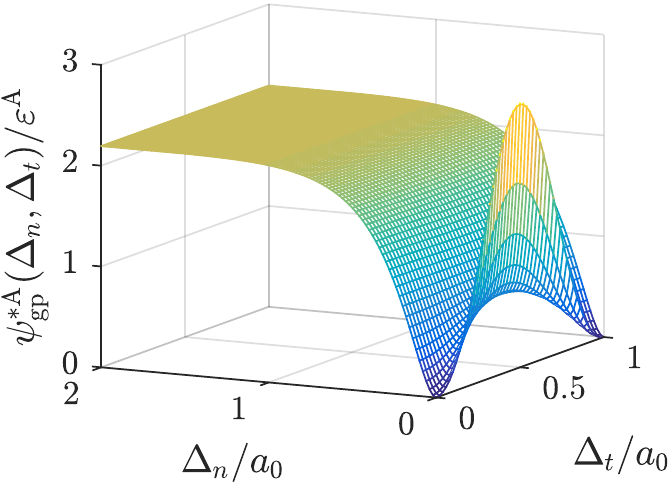}
		\vspace{14pt}
		\label{fig:gsfe}
	\end{subfigure}	
	\caption{(a) Sketch of the applied displacements for the computation of the glide plane potential~$\psi_{\mathrm{gp}}^*(\Delta_n,\Delta_t)$; (b) shape of the glide plane potential corresponding to Phase~A, i.e., $\psi_{\mathrm{gp}}^{*\mathrm{A}}(\Delta_n,\Delta_t)$.}
	\label{fig:msbcs}
\end{figure}

Assuming that the (non-linear) dislocation-related deformation is considered only within the glide plane, the total free energy (per unit out-of-plane thickness) consists of two contributions of the elastic strain energy density~$\psi_{\mathrm{e}}$ (considered inside~$\Omega_\pm^i$) and of the glide plane potential~$\psi_{\mathrm{gp}}$ (non-zero along~$\Gamma_{\mathrm{gp}}^i$):
\begin{equation}
\label{eq:internal-energy}
\Psi=\int_{\Omega\setminus\Gamma_{\mathrm{gp}}}\psi_{\mathrm{e}}\,\mathrm{d}\Omega+\int_{\Gamma_{\mathrm{gp}}}\psi_{\mathrm{gp}}\,\mathrm{d}\Gamma
\end{equation}
Here $\psi_{\mathrm{e}}$ is calculated based on linear elasticity
\begin{equation}
\psi_{\mathrm{e}}=\frac{1}{2}\,\boldsymbol{E}:{}^4\boldsymbol{C}^i:\boldsymbol{E}
\end{equation}	
under a plane strain condition and with the phase specific homogenized fourth-order elasticity tensor ${}^4\boldsymbol{C}^i$. In order to calibrate the PN-FE model to the underlying discrete model, ${}^4\boldsymbol{C}$ is computed from the MS system of Section~\ref{sect:MS} through homogenization following the standard definition as follows \cite[see, e.g.,][]{Tadmor2011}
\begin{equation}
C_{ijkl} = \frac{1}{2V}
\sum_{\substack{\alpha,\beta\\\alpha\neq\beta}}
\left[\phi''(r^{\alpha\beta})-\frac{\phi'(r^{\alpha\beta})}{r^{\alpha\beta}}\right]
\frac{r^{\alpha\beta}_{i} r^{\alpha\beta}_{j} r^{\alpha\beta}_{k} r^{\alpha\beta}_{l}}{(r^{\alpha\beta})^2},
\quad \text{with} \quad
\phi''(r) = \frac{\mathrm{d}^2\phi^{\alpha\beta}(r)}{\mathrm{d}r^2}
\label{eq:stiffness}
\end{equation}
where~$V=A\cdot r_{\mathrm{m}}^{\mathrm{A}}$ denotes the (virtual) volume of the simulation cell in the deformed configuration with the according in-plane sectional area~$A$ (where we have again considered a virtual thickness $r_{\mathrm{m}}^{\mathrm{A}}$). The (isotropic) stiffness tensor is obtained for a doubly periodic simulation cell of dimensions~$20 a_0 \times 12 a_0 \sqrt{3}$. The obtained parameters, computed for Phase~A (cf. Table~\ref{tab:ljsf}), are summarised in Table~\ref{tab:homogenised}. The parameters of Phase~B are obtained by simply multiplying those of Phase~A with the material contrast ratio~$\rho$, i.e., $^4\bs{C}^{\mathrm{B}}=\rho\,^4\bs{C}^{\mathrm{A}}$. The elastic strain tensor is 
\begin{equation}
\boldsymbol{E}=\frac{1}{2}\left(\vec{\nabla}\vec{u} + \left(\vec{\nabla}\vec{u}\right)^T\right)
\end{equation}
The stress tensor follows accordingly as
\begin{equation}
\boldsymbol{\sigma}^i={}^4\boldsymbol{C}^i:\boldsymbol{E}
\end{equation}
%
%
The glide plane potential~$\psi_{\mathrm{gp}}$ is a function of the displacement jump (or disregistry) between~$\Omega^i_+$ and~$\Omega^i_-$ across the glide plane~$\Gamma_{\mathrm{gp}}$, expressed as
\begin{equation}
\label{eq:Disregistry}
\vec{\Delta} = \llbracket \vec{u} \rrbracket=\vec{u}_+-\vec{u}_-
\end{equation} 
To capture the effect of lattice periodicity, $\psi_{\mathrm{gp}}$ is a periodic function in~$\Delta_t$, and to represent the underlying lattice properly, $\psi_{\mathrm{gp}}$ approximates the reference glide plane potential of the atomistic system, $\psi_{\mathrm{gp}}^*$, as close as possible. The different glide plane potentials which are employed in this paper are introduced and calibrated in Section~\ref{sect:GP-models}. Note that the elasticity and glide plane properties are homogeneous in each phase, which results in a jump in material properties across~$\Gamma_{\mathrm{pb}}$, in contrast to the atomistic model in which the non-local interaction between both phases results in a smooth transition. The reference glide plane potential, $\psi_{\mathrm{gp}}^*$, represents the misfit energy between two rigidly shifted bulks of atoms, as shown in Figure~\ref{fig:msbcs_b}. $\psi_{\mathrm{gp}}^*$ is computed by considering a lattice with the stress-free spacing~$a_0$ in a simulation box (again of size~$20 a_0 \times 12 a_0 \sqrt{3}$, recall Eq.~\eqref{eq:stiffness} and the surrounding discussion). The lattice is first equilibrated with periodic boundary conditions applied in the $x$-direction while the top and bottom edges are left free. The glide plane potential is subsequently computed by shifting the upper part of the lattice rigidly according to~$\vec{\Delta} = \Delta_t\vec{e}_t + \Delta_n\vec{e}_n$, where~$\Delta_t$ and~$\Delta_n$ denote the tangential and normal disregistry (relative displacements) across the considered interface. These disregistry components are varied in the ranges $0 \le \Delta_t \le a_0$ and $-0.2\,a_0 \le \Delta_n \le 2\, a_0$. Note that the potential for $\Delta_t < 0$ and $\Delta_t>a_0$ follows by symmetry and periodicity -- in fact, the analysis could have been limited to $0\le\Delta_t\le a_0/2$ for the same reason. For each of these states the energy of the system is evaluated and the energy of the initial stress-free system is subtracted. The obtained result for Phase~$\mathrm{A}$ is shown in Figure~\ref{fig:gsfe}. Reference values of~$\psi_{\mathrm{gp}}^*$ are the work of separation~$G_{\mathrm{c}} = \lim\limits_{\Delta_n\rightarrow\infty}\psi_{\mathrm{gp}}^*$, the unrelaxed unstable stacking fault energy~$\gamma_{\mathrm{us}}^{(\mathrm{u})} = \psi_{\mathrm{gp}}^*(\Delta_n=0,\Delta_t=a_0/2)$, the relaxed unstable stacking fault energy~$\gamma_{\mathrm{us}}^{(\mathrm{r})} = \psi_{\mathrm{gp}}^*(\Delta_n=\Delta_n^\dagger,\Delta_t=a_0/2)$, where~$\Delta_n^\dagger(\Delta_t)$ is the normal disregistry of zero normal traction~$T_n(\Delta_t) = \partial\psi_{\mathrm{gp}}^*(\Delta_t)/\partial\Delta_n = 0$, and the characteristic length $l_{\mathrm{c}}$, i.e., the normal disregistry $\Delta_n$ at $\Delta_t=0$ where $\partial^2\psi_{\mathrm{gp}}^*(\Delta_t)/\partial\Delta_n^2 = 0$. The values obtained for the landscape of Figure~\ref{fig:gsfe} are listed in Table~\ref{tab:gp_potential}. In analogy to the elasticity parameters, the potential of Phase~B is obtained by scaling the potential of Phase~A by the material contrast ratio~$\rho$, i.e., $\psi_{\mathrm{gp}}^{*\mathrm{B}}=\rho\psi_{\mathrm{gp}}^{*\mathrm{A}}$.
\begin{table}
	\caption{Homogenised elasticity parameters corresponding to Phase~$\mathrm{A}$.}
	\renewcommand{\arraystretch}{1.3}
	\centering
	\begin{tabular}{c|c|c|c}
		Parameter & $C_{1111} = C_{2222}$ & $C_{1122}$ & $C_{1212}$ \\\hline \rule{0pt}{2.3ex}
		Value & $102.520 \,\varepsilon^{\mathrm{A}}/{r_{\mathrm{m}}^{\mathrm{A}}}^3$ & $34.173\,\varepsilon^{\mathrm{A}}/{r_{\mathrm{m}}^{\mathrm{A}}}^3$  & $34.173\,\varepsilon^{\mathrm{A}}/{r_{\mathrm{m}}^{\mathrm{A}}}^3$ 
	\end{tabular}
	\label{tab:homogenised}
\end{table}
\begin{table}
	\caption{Reference values of the glide plane potential of Phase~$\mathrm{A}$.}
	\renewcommand{\arraystretch}{1.3}
	\centering
	\begin{tabular}{c|c|c|c|c|c}
		Parameter & $G_{\mathrm{c}} $ & $\gamma_{\mathrm{us}}^{(\mathrm{u})}$ & $\gamma_{\mathrm{us}}^{(\mathrm{r})}$ & $l_{\mathrm{c}}$ & $\Delta_n^\dagger(\Delta_t=a_0/2)$ \\\hline \rule{0pt}{2.3ex}
		Value & $2.19\,\varepsilon^{\mathrm{A}}/{r_{\mathrm{m}}^{\mathrm{A}}}^2$ & $2.61\,\varepsilon^{\mathrm{A}}/{r_{\mathrm{m}}^{\mathrm{A}}}^2$ & $0.71\,\varepsilon^{\mathrm{A}}/{r_{\mathrm{m}}^{\mathrm{A}}}^2$ & $0.127\,a_0$ & $0.119\,a_0$
	\end{tabular}
	\label{tab:gp_potential}
\end{table}
\par
%
%
The perfectly bonded and fully coherent phase boundary~$\Gamma_{\mathrm{pb}}$ is modelled by enforcing displacement and traction continuity, i.e.,
\begin{alignat}{3}
\vec{u}^{\mathrm{A}} &= \vec{u}^{\mathrm{B}}, & \quad\mathrm{on}\:\Gamma_{\mathrm{pb}}\label{eq:pb-traction}\\
\boldsymbol{\sigma}^{\mathrm{A}}\cdot\vec{e}_x &= \boldsymbol{\sigma}^{\mathrm{B}}\cdot\vec{e}_x, & \quad\mathrm{on}\:\Gamma_{\mathrm{pb}}\label{eq:pb-displacement}
\end{alignat}
\par
Similarly to the atomistic model, the shear deformation is applied on the external boundary~$\partial\Omega$ in the form of Dirichlet boundary conditions following Eqs.~\eqref{eq:bcs-a} and~\eqref{eq:bcs-b}. Under such constraints, the mechanical equilibrium inside~$\Omega$ is established by minimising the total potential energy~$\Psi$ of Eq.~\eqref{eq:internal-energy} subject to the phase boundary constraints of Eqs.~\eqref{eq:pb-traction} and~\eqref{eq:pb-displacement}. Due to the periodicity of~$\psi_{\mathrm{gp}}$ this formulation results in a non-convex minimisation problem, which is solved using the finite element method by discretising~$\Omega_\pm^i$ and~$\Gamma_{\mathrm{gp}}^i$ in space using linear triangular elements for the bulk and linear line elements for the glide plane. The largest element size on the glide plane~$\Gamma_{\mathrm{gp}}$ as well as on both phase boundaries~$\Gamma_{\mathrm{pb}}$ is limited to~$a_0/4$. This size is based on a thorough mesh convergence study \cite[reported in][Section~4]{Bormann2018}, and is sufficient to yield a negligible Peierls barrier \cite[see, e.g.,][]{Xu:2020}. The resulting total potential energy is minimised with the help of the truncated Newton optimization algorithm, see, e.g., \cite{Bormann2018b}. The two algorithms used for the minimization of the MS and PN-FE energies, i.e. Trust-region for Eq.~\eqref{eq:ms_min} and truncated Newton for Eq.~\eqref{eq:internal-energy}, have been chosen because of their good performance, fast convergence, and robustness for the particular problem considered. Since both the MS and PN-FE seek for the closest local minima in non-convex energy landscapes, as long as any possible alternative solvers which could be used instead are robust enough and initialized well enough, the results and comparison presented below in Section~\ref{section:Comparison} should not be affected. Individual dislocations are initialized using an approach identical to the one described at the end of Section~\ref{sect:MS}, where the critical nucleation stress~$\tau_{\mathrm{nuc}} = 0.012\frac{\mu^{\mathrm{A}}}{2\pi}$ is used~\citep{Bormann2018}, and where the Virial stresses~$\tau(x)$ are replaced with the resolved shear stresses acting on the glide plane~$\Gamma_{\mathrm{gp}}$ (as obtained from the FE simulation).
%
%
\section{Glide plane models and their calibration}
\label{sect:GP-models}
To study the influence of the glide plane potential~$\psi_{\mathrm{gp}}$ on the accuracy of the PN-FE framework outlined in Section~\ref{sect:pnfe}, different complexities of~$\psi_{\mathrm{gp}}$ are considered in this section. The calibration of each potential is discussed for Phase~$\mathrm{A}$ only; the properties of the Phase~$\mathrm{B}$ potential follow accordingly using the material contrast ratio $\rho$. First, the classical sinusoidal PN potential, which depends only on the tangential opening~$\Delta_t$, is described in Section~\ref{sect:pot:pnfe}. The accuracy of this model can be improved by considering a Fourier series, as described in Section~\ref{sect:pot:fourier}. A coupling of the tangential~$\Delta_t$ as well as the normal~$\Delta_n$ disregistry is introduced through an analytical model outlined in Section~\ref{sect:pot:coupledana}, whereas the direct use of the numerical data obtained from the MS calculations (i.e., $\psi_{\mathrm{gp}}^*(\Delta_n,\Delta_t)$, recall Figure~\ref{fig:gsfe}) is described in Section~\ref{sect:pot:couplednum}. 
%
%
\subsection{Classical Peierls--Nabarro potential}
\label{sect:pot:pnfe}
The classical assumption for the glide plane potential is based on the Frenkel sinusoidal model, expressed as a function of the tangential disregistry~$\Delta_t$ as
\begin{equation}
\label{eq:PN-classic}
\psi_{\mathrm{PN}} = \gamma_{\mathrm{us}}\sin^2\left(\frac{\pi\Delta_t}{b}\right)
\end{equation} 
where $\gamma_{\mathrm{us}}$ is the unstable stacking fault energy and~$b$ the Burgers vector magnitude. The normal disregistry is constrained to zero, i.e., $\Delta_n = 0$. To calibrate this potential to the MS simulations, $b$ is chosen to be equal to the atomic spacing~$a_0$ to match the lattice periodicity. The unstable stacking fault energy, $\gamma_{\mathrm{us}}$, is determined from the atomistically calculated energy landscape~$\psi_{\mathrm{gp}}^*(\Delta_n,\Delta_t)$, recall Section \ref{sect:pnfe}. Keeping in mind that the system evolves towards a state of minimum energy, two potentials are considered to investigate the influence of the value of~$\gamma_{\mathrm{us}}$. The first potential, $\psi_{\mathrm{PN},\mathrm{r}}$, is calibrated by setting~$\gamma_{\mathrm{us}}$ equal to the relaxed unstable stacking fault energy of~$\psi_{\mathrm{gp}}^*$, i.e., $\gamma_{\mathrm{us}} = \gamma_{\mathrm{us}}^{(\mathrm{r})}$. Note that this choice introduces a slight inconsistency with the constraint $\Delta_n=0$. The second potential, $\psi_{\mathrm{PN},\mathrm{u}}$, employs~$\gamma_{\mathrm{us}} = \gamma_{\mathrm{us}}^{(\mathrm{u})}$ instead, i.e., unstable unrelaxed stacking fault energy. Both potentials~$\psi_{\mathrm{PN},i}$ along with their tractions~$T_{t,i}=\partial\psi_{\mathrm{PN},i}/\partial\Delta_t$, $i\in\{\mathrm{r},\mathrm{u}\}$, are shown in Figure~\ref{fig:Energy_Traction_GP} as a function of the tangential disregistry~$\Delta_t$.
%
%
\subsection{Fourier potential}
\label{sect:pot:fourier}
Assuming that for any~$\Delta_t$ the glide plane potential always occupies the state of minimum energy in~$\psi_{\mathrm{gp}}^*$ for that $\Delta_t$, i.e., where $T_n^*(\Delta_t) = \partial\psi_{\mathrm{gp}}^*/\partial\Delta_n=0$, the corresponding potential~$\psi_{\mathrm{gp}}(\Delta_t)$ cannot be represented with sufficient accuracy by the simple sinusoidal function of Eq.~\eqref{eq:PN-classic}. Although the amplitudes of the energy and traction are captured accurately, their profiles are inaccurate (cf. Figure~\ref{fig:Energy_Traction_GP}, in which the dashed curves represent the atomistic data). A higher complexity is therefore required, which can be achieved through a Fourier series expansion for the glide plane tractions. The corresponding potential is obtained through integration and reads
\begin{equation}
\label{eq:PN-Fourier}
\psi_{\mathrm{F}}=\sum_{k=1}^{n}\frac{1}{k}\gamma_{\mathrm{us},k}\sin^2\left(\frac{k\pi\Delta_t}{b}\right)
\end{equation}
The Fourier coefficients~$\gamma_{\mathrm{us},k}$ in Eq.~\eqref{eq:PN-Fourier} are determined by a least-squares fit of the shear traction~$T_{t,\mathrm{F}} = \partial\psi_{\mathrm{F}}/\partial\Delta_t$ to~$T_t^*(\Delta_n^\dagger,\Delta_t)$. In analogy to the classical PN potential, the normal disregistry is neglected. The first four obtained Fourier coefficients are listed in Table~\ref{tab:fourier_parameters}, whereas higher order terms (i.e., $k>4$) are dropped because of their negligible influence. The fitted potential~$\psi_{\mathrm{F}}(\Delta_t)$ and its traction~$T_{t,\mathrm{F}}(\Delta_t)$ are shown in Figure~\ref{fig:Energy_Traction_GP}. As can be observed, it fits the atomistic data nearly perfectly along the relaxed path ($\Delta_n^\dagger, \Delta_t$) in the diagrams. 
\begin{table}[htbp]
	\caption{Fitted parameters for the Fourier glide plane potential~$\psi_{\mathrm{F}}$.}
	\renewcommand{\arraystretch}{1.2}
	\centering
	\begin{tabular}{c|c|c|c|c|c}
		Parameter & $\gamma_{\mathrm{us},1}$ & $\gamma_{\mathrm{us},2}$ & $\gamma_{\mathrm{us},3}$ & $\gamma_{\mathrm{us},4}$ & $b$\\\hline \rule{0pt}{2.3ex}
		Value & $0.667 \,\varepsilon^{\mathrm{A}}/r_{\mathrm{m}}^2$ & $0.296\,\varepsilon^{\mathrm{A}}/r_{\mathrm{m}}^2$ & $0.112\,\varepsilon^{\mathrm{A}}/r_{\mathrm{m}}^2$ & $0.039\,\varepsilon^{\mathrm{A}}/r_{\mathrm{m}}^2$ & $a_0$\\
	\end{tabular}
	\label{tab:fourier_parameters}
\end{table}
%
%
\subsection{Coupled analytical potential}
\label{sect:pot:coupledana}
Instead of restricting the potential~$\psi_{\mathrm{gp}}$ to represent only one section~$\psi_{\mathrm{gp}}^*(\Delta_n^\dagger,\Delta_t)$, the full potential~$\psi_{\mathrm{gp}}^*(\Delta_n, \Delta_t)$ can be approximated by introducing an additional dependency of~$\psi_{\mathrm{gp}}$ on the normal disregistry~$\Delta_n$. As a result, also the out-of-plane reaction force of the glide plane is included, as opposed to the previous models which neglected this contribution. Following \cite{Sun1993}, the coupled potential can be expressed as
\begin{align}
\label{eq:Sun}
\psi_{\mathrm{C}}(\Delta_n, \Delta_t) 
&= G_{\mathrm{c}}\left\{1-\left[1+\frac{\Delta_n}{l_c}\right]\exp\left(-\frac{\Delta_n}{l_c}\right)\right.\nonumber\\
& \left.+\sin^2\left(\frac{\pi\Delta_t}{b}\right)\left[q+\frac{q-p}{1-p}\left(\frac{\Delta_n}{l_c}\right)\right]\exp\left(-\frac{\Delta_n}{l_c}\right)\right\}
\end{align}
with the work of separation~$G_{\mathrm{c}}$, the characteristic length~$l_c$ (i.e., such a normal disregistry~$\Delta_n$ at~$\Delta_t=0$ for which~$\partial^2\psi_{\mathrm{gp}}^*/\partial\Delta_n^2 = 0$, recall Table~\ref{tab:gp_potential}), and~$p=\Delta_n^\dagger(\Delta_t=b/2)/l_c$. \cite{Sun1993} further suggested to set~$q$ equal to the ratio~$\gamma_{\mathrm{us}}^{(\mathrm{u})}/G_{\mathrm{c}}$. However, this, along with Eq.~\eqref{eq:Sun}, leads to a strong overestimation of~$\gamma_{\mathrm{us}}^{(\mathrm{r})}$ as~$\psi_{C}(\Delta_n^\dagger,b/2)\approx 7\gamma_{\mathrm{us}}^{(\mathrm{r})}$. Considering~$\gamma_{\mathrm{us}}^{(\mathrm{r})}$ instead of~$\gamma_{\mathrm{us}}^{(\mathrm{u})}$ to be a key quantity of the glide plane potential, $q$ is to be calibrated such that~$\gamma_{\mathrm{us}}^{(\mathrm{r})}$ is represented correctly. This relationship yields, cf.~\citep{Sun1993},
\begin{equation}
q = \left(\frac{\gamma_{\mathrm{us}}^{(\mathrm{r})}}{G_{\mathrm{c}}}-1\right)\left(1-p\right)\exp(p)+1
\end{equation}
The correspondingly calculated parameters for~$\psi_{\mathrm{C}}$ are listed in Table~\ref{tab:coupled_parameters}. The energy~$\psi_{\mathrm{C}}$ and shear traction~$T_{t,\mathrm{C}}$ are plotted in Figure~\ref{fig:Energy_Traction_GP} along~$\Delta_n^\dagger$, $\Delta_t$, to allow for an adequate comparison with the other glide plane potentials. 
\begin{table}[htbp]
	\caption{Fitted parameters for the coupled analytical glide plane potential~$\psi_{\mathrm{C}}$.}
	\renewcommand{\arraystretch}{1.2}
	\centering
	\begin{tabular}{c|c|c|c|c|c}
		Parameter & $G_{\mathrm{c}}$ & $l_c$ & $p$ & $q$ & $b$\\\hline \rule{0pt}{2.3ex}
		Value & $2.194\,\varepsilon^{\mathrm{A}}/r_{\mathrm{m}}^2$ & $0.127\,a_0$ & $0.9357$ & $0.8887$ & $a_0$
	\end{tabular}
	\label{tab:coupled_parameters}
\end{table}
%
%
\subsection{Coupled numerical potential}
\label{sect:pot:couplednum}
\begin{figure}
	\centering
	\begin{subfigure}{0.49\textwidth}{\label{first}}
		\centering
		\includegraphics[width=1.\textwidth]{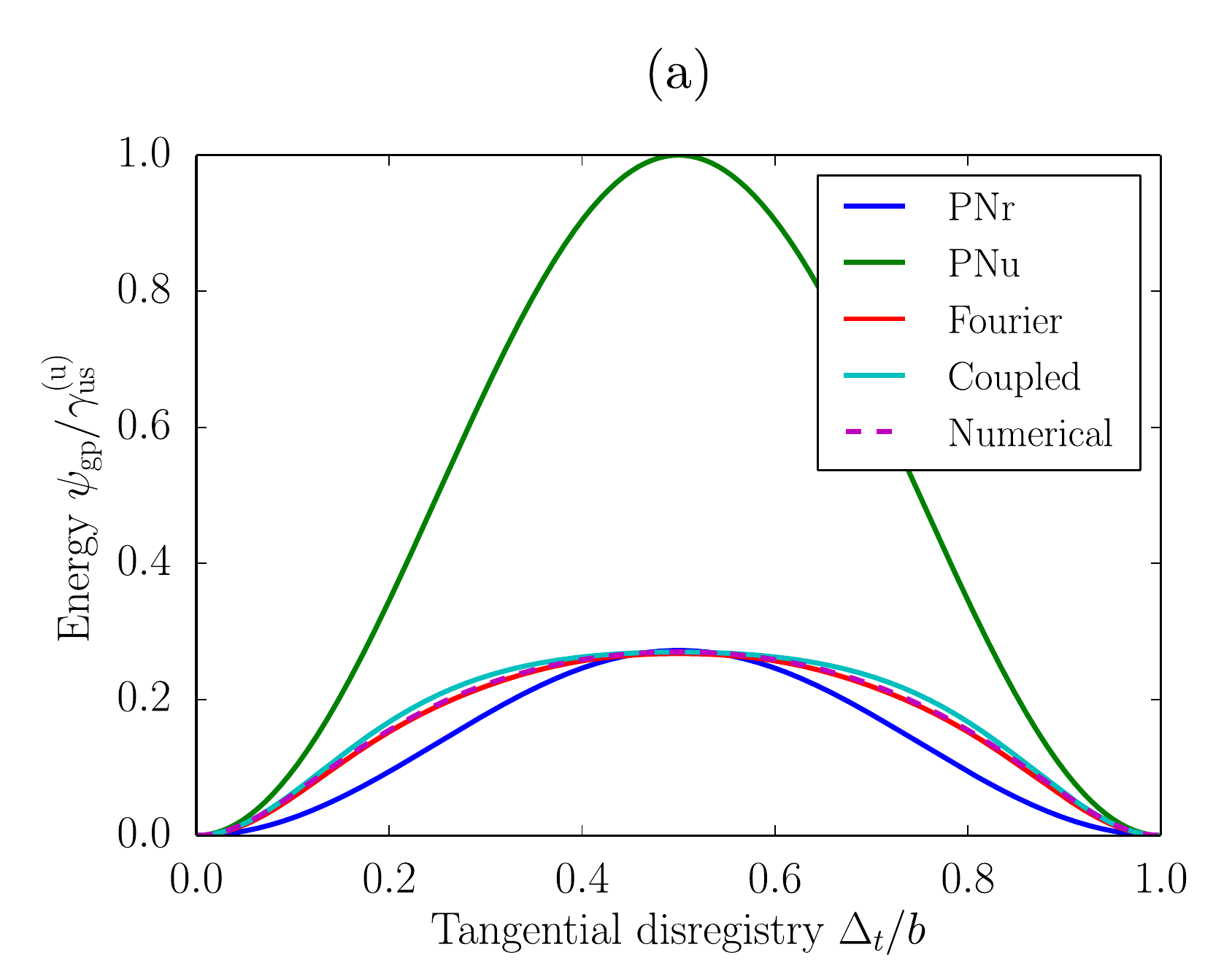}
		\label{fig:Energy_GP}
	\end{subfigure}	
	\begin{subfigure}{0.49\textwidth}
		\centering
		\includegraphics[width=1.\textwidth]{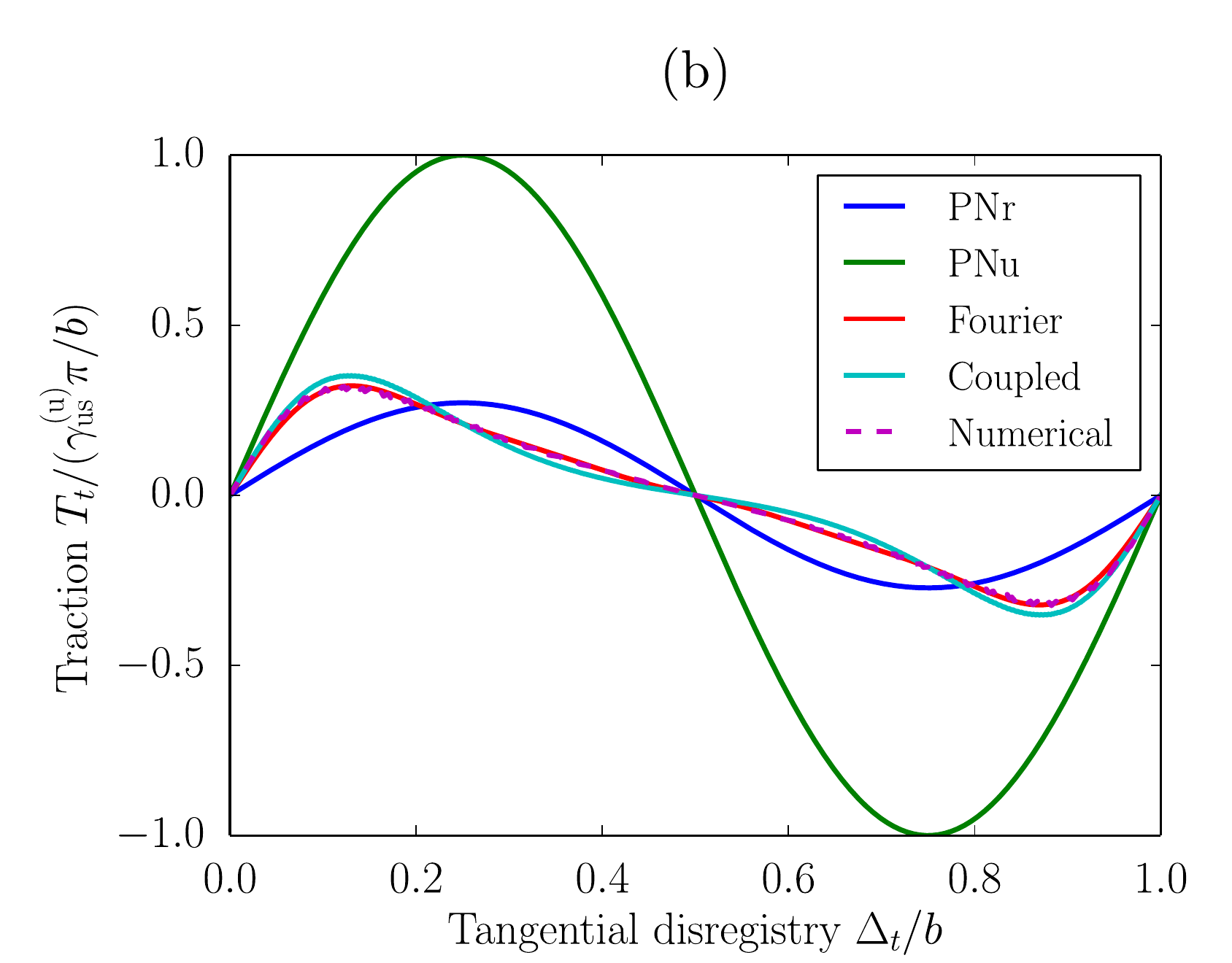}
		\label{fig:Traction_GP}
	\end{subfigure}
	\caption{Various employed glide plane models, plotted along~$\Delta_n^\dagger$, $\Delta_t$, where~$\Delta_n^\dagger(\Delta_t)$ is the normal disregistry~$\Delta_n$ for which the normal traction vanishes, i.e., $T_n = 0$: (a) glide plane potential $\psi_{\mathrm{gp}}$; (b) glide plane shear traction~$T_t$.}
	\label{fig:Energy_Traction_GP}
\end{figure}

The highest possible accuracy of the glide plane potential for the PN-FE model is obtained if the data points of~$\psi_{\mathrm{gp}}^*$ are directly used as input for the glide plane potential, which is denoted as~$\psi_{\mathrm{N}}$. Intermediate values are calculated numerically via a cubic interpolation scheme to ensure sufficient smoothness of the glide plane traction~$\partial\psi_{\mathrm{N}}/\partial\vec{\Delta}$ and stiffness~$\partial^2\psi_{\mathrm{N}}/\partial\vec{\Delta}^2$. Figure~\ref{fig:Energy_Traction_GP} shows the energy~$\psi_{\mathrm{N}}$ and corresponding traction~$T_{t,\mathrm{N}}$ along~$\Delta_n^\dagger$, $\Delta_t$. 
%
%
\section{Comparative analysis}
\label{section:Comparison}
In this section a two-fold objective is pursued. (i) The predictive capability of the PN-FE methodology is assessed by a comparison against the MS simulation. To ensure the optimum accuracy of the PN-FE results, the coupled numerical glide plane potential~$\psi_{\mathrm{N}}(\Delta_n, \Delta_t)$ (cf. Section~\ref{sect:GP-models}) is used. (ii) The influence of the various glide plane potentials (cf. Section \ref{sect:GP-models}) on the obtained results is studied. \par
First, the employed geometric properties of the two-phase microstructure are specified in Section~\ref{sect:parameters}. The results for a single dislocation are discussed in Section~\ref{sect:single_disl} in terms of the disregistry and traction profiles, as well as the displacement and strain fields. Similarly, the results for a three-dislocation pile-up configuration follow in Section~\ref{sect:disl-pu}. Finally, the full evolution process, from zero dislocations to dislocation transmission, is discussed in Section~\ref{sect:dislocations} by means of the dislocation positions and the total free energy.\par
%
%
\subsection{Parameter set}
\label{sect:parameters}
The two-phase microstructure, as introduced in Section~\ref{section:Problem_Statement}, is considered, with~$512\times 516$ atoms in Phase~A and two times~$258\times 516$ atoms in Phase~B. After subtracting the rigid boundary layer~$\Omega^{\mathrm{BC}}$, the model dimensions of the PN-FE model are~$L\times H = 511.5\,a_0\times 256\sqrt{3}\, a_0$ and~$L_{\mathrm{A}}=256\,a_0$. The phase contrast is chosen as~$\rho=1.4$ and the constant lattice spacing implies~$b^{\mathrm{B}}=b^{\mathrm{A}}=b$. The critical nucleation stress for the PN-FE model is set to~$\tau_{\mathrm{nuc}} = 0.0035\,\mu^{\mathrm{A}}$. Due to the symmetry of the problem ($\vec{u}(-\vec{x})=-\vec{u}(\vec{x})$), the atomistic results are plotted only for~$x \ge 0$. The PN-FE simulations, on the contrary, exploit this symmetry condition and require thus only the consideration of the half-domain~$x\ge 0$. In the following, the atomistic simulation results are labelled as MS. All results are normalised with respect to~$b$ (positions, disregistry), $\mu^{\mathrm{A}}$ (tractions, stresses), and~$\varepsilon^{\mathrm{A}}$ (energy).

Recall that the investigated system involves a finite specimen (cf. Fig.~\ref{fig:Model} and Section~\ref{sect:MS}), which is studied here on purpose. The results are therefore in multiple ways system size dependent~-- various (mostly smaller) sizes have been tested to make sure that the required behaviour is observed (i.e.~a pile-up containing multiple dislocations, followed by dislocation transmission in later stages of the loading). The absolute system sizes in the MS and the PN-FE models were chosen to be exactly the same, thus allowing for their direct comparison.
%
%
\subsection{Single dislocation}
\label{sect:single_disl}
\subsubsection{Disregistry profile}
\begin{figure}
	\centering
	\begin{subfigure}{0.49\textwidth}
		\centering
		\caption{}
		\includegraphics[width=1.\textwidth]{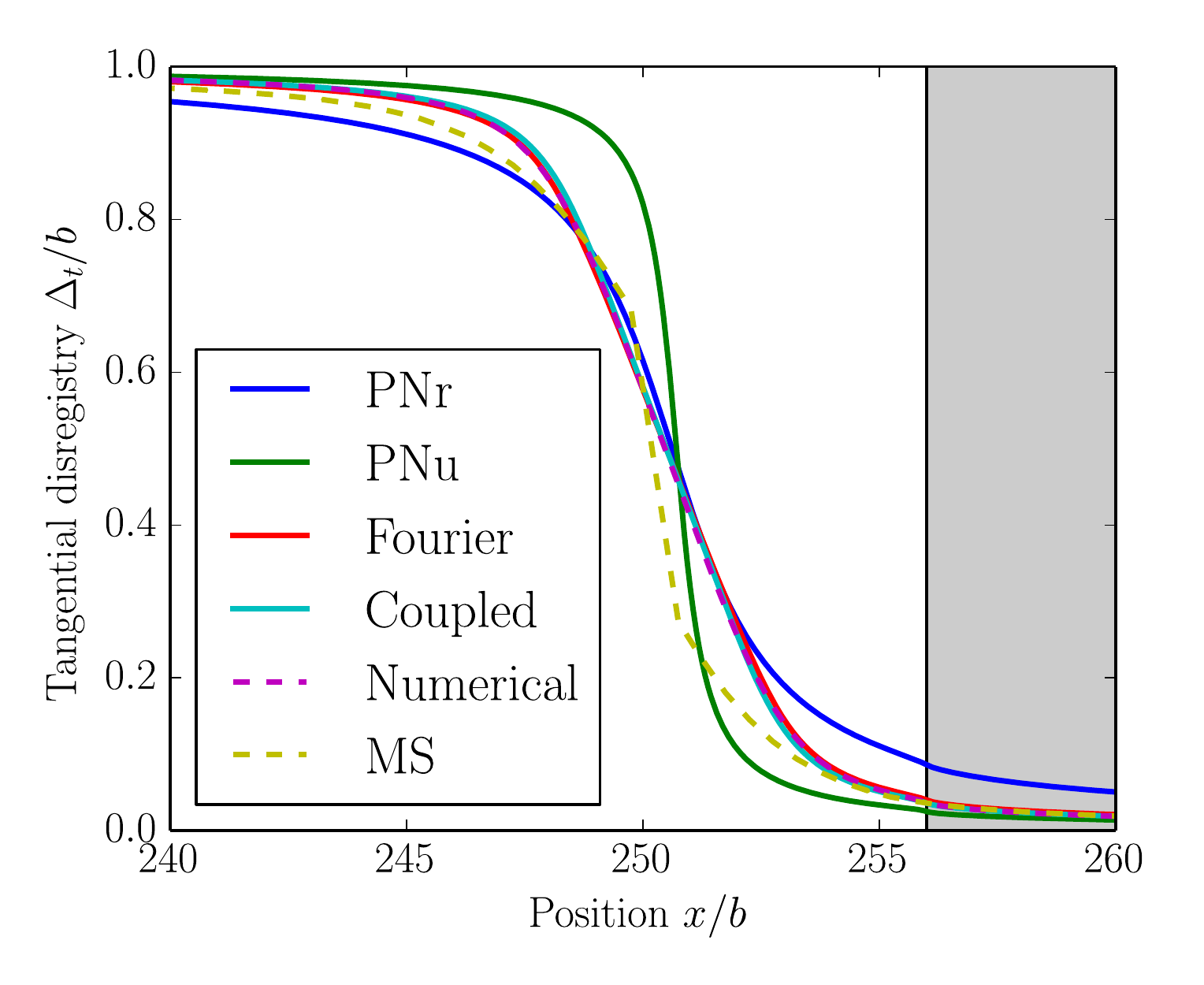}
		\label{fig:Disregistry-tan-step-71}
	\end{subfigure}
	\begin{subfigure}{0.49\textwidth}
		\centering
		\caption{}
		\includegraphics[width=1.\textwidth]{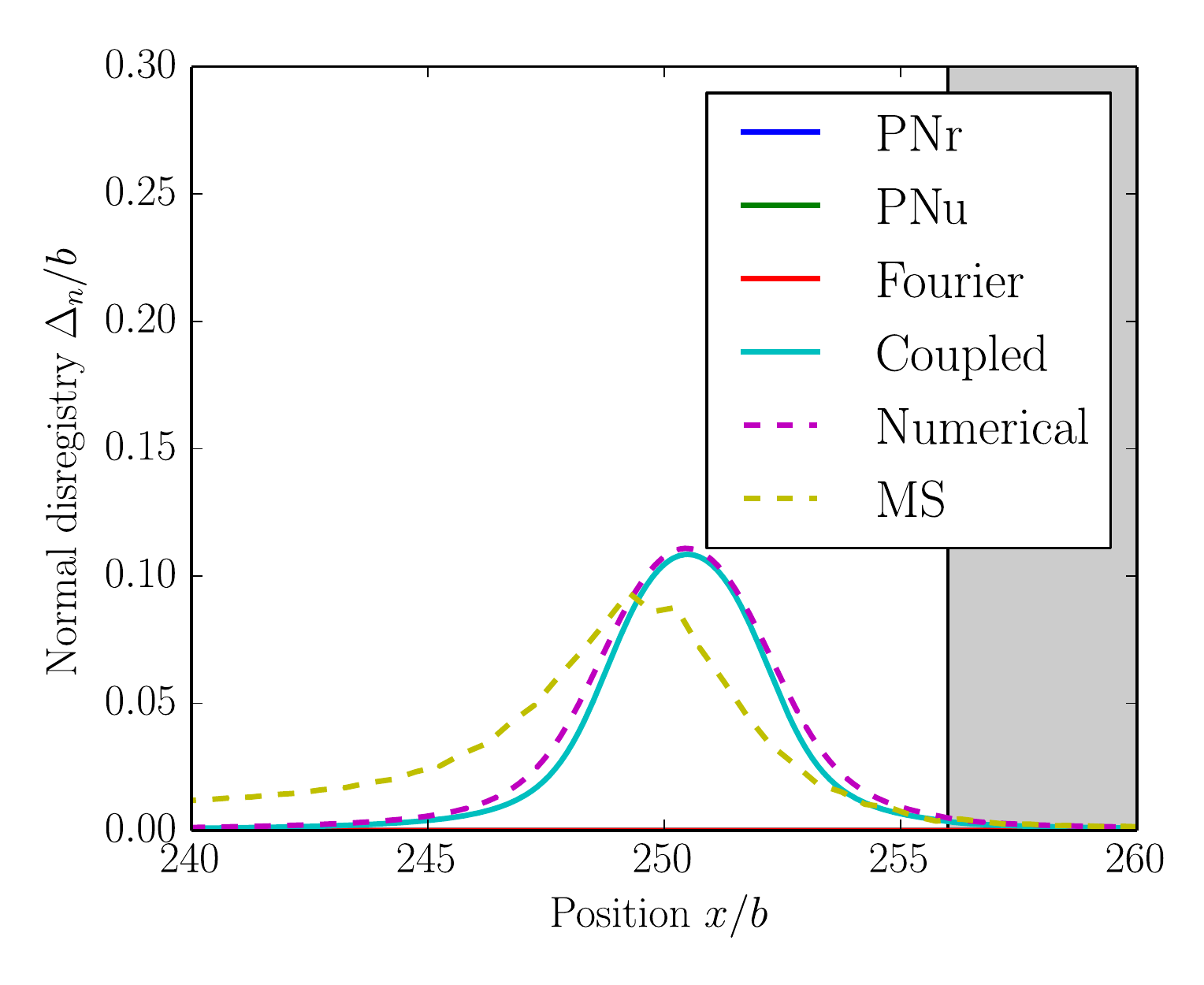}
		\label{fig:Disregistry-norm-step-71}
	\end{subfigure}
	\caption{Disregistry profiles of the different glide plane potentials in comparison with the atomistic model (MS) at~$\tau=0.00473\,\mu^{\mathrm{A}}$: (a) tangential disregistry~$\Delta_t$; (b) normal disregistry~$\Delta_n$. The grey area indicates Phase~B}
	\label{fig:Disregistry-step-71}
\end{figure}

Consider first an externally applied shear load $\tau=0.00473\,\mu^{\mathrm{A}}$, at which in both models (MS and PN-FE) a single dislocation exists, obstructed by the phase boundary. The local dislocation configuration of the MS model and the PN-FE model with its different glide plane potentials is illustrated in Figure \ref{fig:Disregistry-step-71} in terms of the disregistry profiles $\Delta_t$ (Figure \ref{fig:Disregistry-tan-step-71}) and $\Delta_n$ (Figure \ref{fig:Disregistry-norm-step-71}). While for the PN-FE model the disregistries are straightforwardly calculated from the nodal displacements through Eq.~\eqref{eq:Disregistry}, the atomistic model requires first the interpolation of the displacements of atoms above and below the glide plane before Eq.~\eqref{eq:Disregistry} can be applied.\par
As a characteristic of the PN model, the presence of the dislocation is indicated by the drop in disregistry from $\Delta_t=b$ (fully slipped) to $\Delta_t=0$ (non-slipped), which in the PN-FE model is established through the minimisation of the total free energy of Eq. \eqref{eq:internal-energy} -- without the requirement of additional criteria. Naturally, the dislocation is taken to be situated at $\Delta_t=b/2$. \par
The comparison of the PN-FE model (with the coupled numerical glide plane potential $\psi_{\mathrm{N}}$) with the MS model shows that, despite a deviation within the dislocation core, the rather simple PN-FE model is able to approximate the disregistry profiles ($\Delta_t$ and $\Delta_n$) and the dislocation position of the atomistic model reasonably well. The deviations within the dislocation core originate from the local and linear-elastic continuum formulation used in the PN-FE model. Note that the relatively large difference in~$\Delta_n$ for MS and PN-FE for all glide plane potentials in Fig.~\ref{fig:Disregistry-norm-step-71} observed for~$x/b < 247$ is an artefact. The artefact originates from the MS model where $\vec{\Delta}$ is calculated in the reference configuration (at~$t=0$). In this context, atoms which are increasingly separated horizontally by $\Delta_t$ experience in relation with the external shear load an increasing difference in vertical displacement, that leads to an increase in $\Delta_n$.\par
The study of the different analytical glide plane potentials~$\psi_{\mathrm{gp}}$ unveils a significant influence on the local tangential dislocation disregistry profile $\Delta_t$. Consider the PN-FE model with the coupled numerical glide plane potential~$\psi_{\mathrm{N}}$ as the reference solution. The classical PN potentials ($\psi_{\mathrm{PN},\mathrm{r}}$ and~$\psi_{\mathrm{PN},\mathrm{u}}$) exhibit certain deviations in~$\Delta_t$, which can be related to the difference in~$\psi_{\mathrm{gp}}$ as follows (cf. also Figure~\ref{fig:Energy_Traction_GP}a). Using~$\psi_{\mathrm{PN},\mathrm{r}}$ small deviations of~$\Delta_t$ from the zero-energy state ($\Delta_t = i\,b$ with~$i=0,1,2,\dots$) are energetically less penalised, as compared to~$\psi_{\mathrm{N}}$. Consequently, the surrounding bulk outside of the dislocation core relaxes more to reduce elastic strain energy, leading to the slow decay/increase towards~$\Delta_t = i\,b$. On the contrary, $\psi_{\mathrm{PN},\mathrm{u}}$ penalises any tangential disregistry~$\Delta_t$ to a significantly larger extent. The dislocation core is therefore more compressed to reduce the contribution of the glide plane potential to the total energy, resulting not only in a fast decay/increase towards~$\Delta_t = i\,b$, but also in a higher gradient within the dislocation core. Comparing the glide plane potentials~$\psi_{\mathrm{F}}$ and~$\psi_{\mathrm{C}}$ with~$\psi_{\mathrm{N}}$ shows very closely matching tangential disregistry profiles. This supports the assumptions stated in section \ref{sect:GP-models} that: (i)~in the PN-FE model, $\gamma_{\mathrm{us}}^{(\mathrm{r})}$ is the key quantity for the proper description of the dislocation behaviour; (ii)~the disregistry profile~$\vec{\Delta}$ of a dislocation follows the path of minimum energy in~$\psi_{\mathrm{gp}}^*$, i.e., along $\Delta_n^\dagger$. Only the coupled numerical and coupled analytical glide plane potentials~$\psi_{\mathrm{N}}$ and~$\psi_{\mathrm{C}}$ include the dependency on~$\Delta_n$, and are thus capable of approximating the normal disregistry~$\Delta_n$ of the atomistic model. The Fourier potential, however, while neglecting the normal disregistry, nevertheless does a good job at capturing the tangential disregistry by virtue of its calibration on the relaxed energy $\psi_{\mathrm{gp}}^*$. As a possible alternative to the Fourier potential~$\psi_{\mathrm{F}}$ of Section~\ref{sect:pot:fourier}, a two-dimensional glide plane Fourier potential reported by \cite{Schoeck2001} could be used, which depends on the tangential~$\Delta_t$ as well as normal~$\Delta_n$ disregistry, i.e.~$\psi_{\mathrm{F}}(\Delta_n, \Delta_t)$. From the results presented above it may be clear that the two-dimensional version would not bring much added value. The reason for this is that the numerical potential~$\psi_{\mathrm{N}}(\Delta_n, \Delta_t)$, has the highest possible accuracy one can possibly reach (including the full dependence on~$\Delta_n$), and yet in comparison with the Fourier potential~$\psi_{\mathrm{F}}$ (completely omitting the dependence on~$\Delta_n$) does not seem to be more accurate. The absence of the~$\Delta_n$ influence is thus not the limiting factor for the adopted Fourier potential.\par
In general all considered glide plane potentials of the PN-FE model achieve good accuracy in the dislocation core size. This observation appears to be inconsistent with the literature, see, e.g., \cite{Xu:2019b}, where the core disregistries tend to be too narrow for the PN-FE model as compared to MS. A possible explanation is the difference between 2D versus 3D simulations: whereas the work of~\cite{Xu:2019b} uses full 3D simulations, the results presented above are confined to 2D approximations. Because in 3D simulations the dislocation is allowed to relax into the third (out-of-plane) direction, the behaviour of the two models might slightly differ in terms of dislocation core widths.\par
\subsubsection{Glide plane tractions}
\label{sect:tractions}
\begin{figure}[htbp]
	\centering
	\centering
	\includegraphics[width=0.5\textwidth]{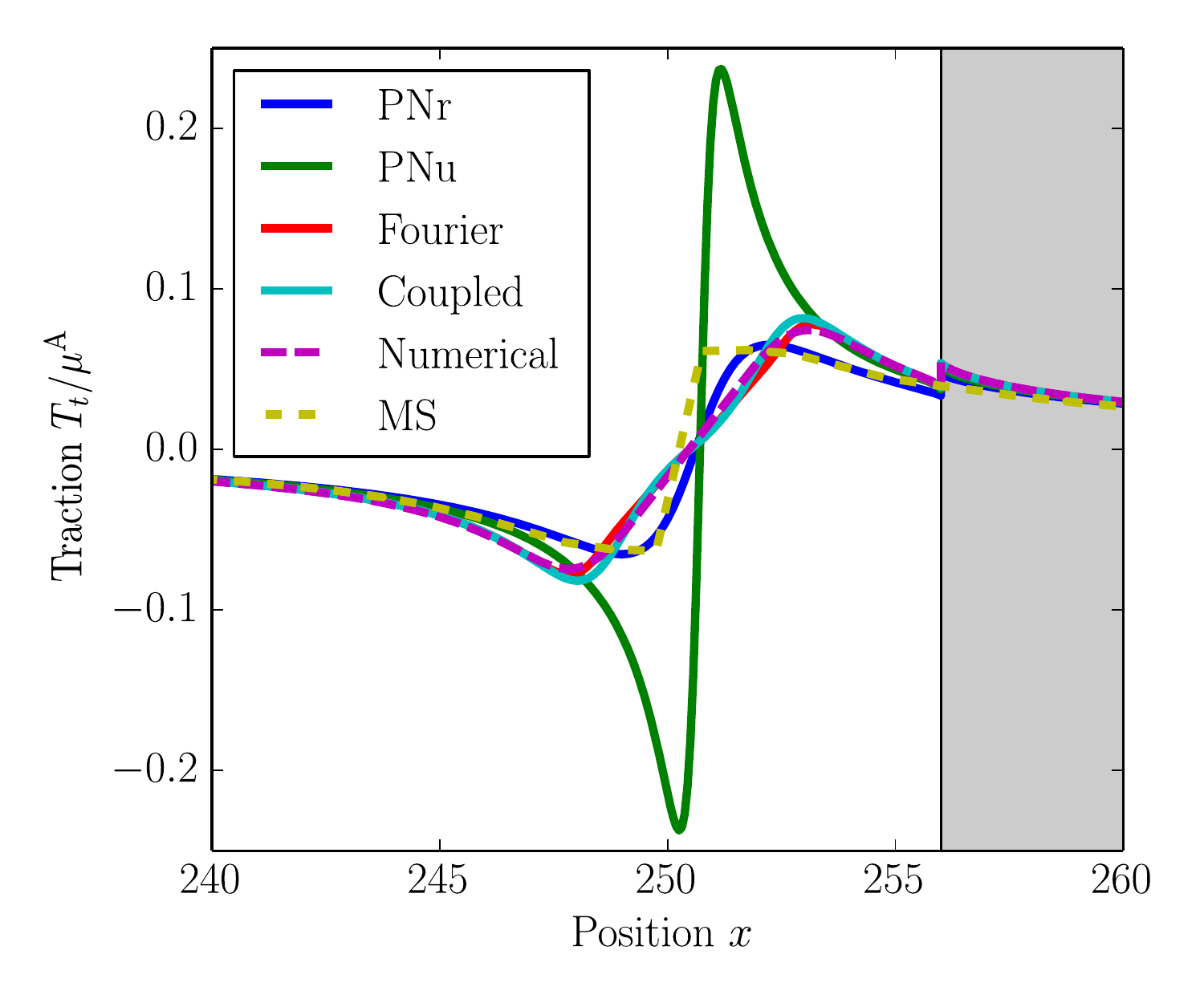}
	\caption{Glide plane shear traction~$T_t$ corresponding to the PN-FE model with the different glide plane potentials and to the atomistic model (MS), at external loading corresponding to~$\tau=0.00473\,\mu^{\mathrm{A}}$. The grey area indicates Phase~B.}
	\label{fig:Shear-Traction-step-71}
\end{figure}

We next discuss the predictions made with the PN-FE model in terms of the dislocation-induced glide plane tractions $T_t$, and their implications for the dislocation behaviour. A comparison of the tractions is presented in Figure~\ref{fig:Shear-Traction-step-71}, corresponding to the disregistry profiles in Figure~\ref{fig:Disregistry-step-71}. The atomistic tractions are taken as the pointwise Virial shear stress~$\sigma_{xy}$ (recall Eq.~\eqref{eq:Virialstress}) at atom positions below the glide plane. Note that the Virial stress, since it is an averaged quantity over neighbouring atoms, becomes questionable within the dislocation core. Hence, no comparison between the atomistic model and the PN-FE model in the core region ($\pm 5b$) follows. Outside of the dislocation core, nevertheless, it can be seen that the traction distribution in the PN-FE model (with $\psi_{\mathrm{N}}$) agrees to the MS model. A minor irregularity of the traction, a discontinuity at the phase boundary, is present in the PN-FE model, which relates to the discontinuity in material properties across the phase boundary. The similarity outside of the dislocation core exemplifies the good representative capability of the PN-FE model for dislocation-induced long-range stresses in two-phase microstructures. For sufficiently separated dislocations in a pile-up ($>5b$), similar dislocation-induced repulsive shear stresses can thus be expected.\par
A comparison of the glide plane potentials in terms of the shear tractions $T_t$ shows no influence on the tractions outside of the dislocation core ($\pm 5b$). Hence, no difference in the dislocation-dislocation interaction is expected if dislocations are no closer than $\approx 5b$. Within the dislocation core, however, and similar to the tangential disregistry profile, a strong influence on the tractions is unveiled. Thus, different repulsive shear stresses occur for sufficiently close dislocations (i.e., $\le 5b$). In such a circumstance the classical PN potentials would exhibit a substantially larger (PNu with $\gamma_{\mathrm{us}}^{(\mathrm{u})}$) or slightly lower (PNr with $\gamma_{\mathrm{us}}^{(\mathrm{r})}$) repulsion as compared to the coupled numerical potential. The Fourier potential and the coupled analytical potential, on the contrary, are invoking a similar repulsion as the coupled numerical potential.\par
\subsubsection{Displacement and strain fields}
\label{sect:single_disl_fields}
\begin{figure}[htbp]
	\centering
	\includegraphics[scale=1]{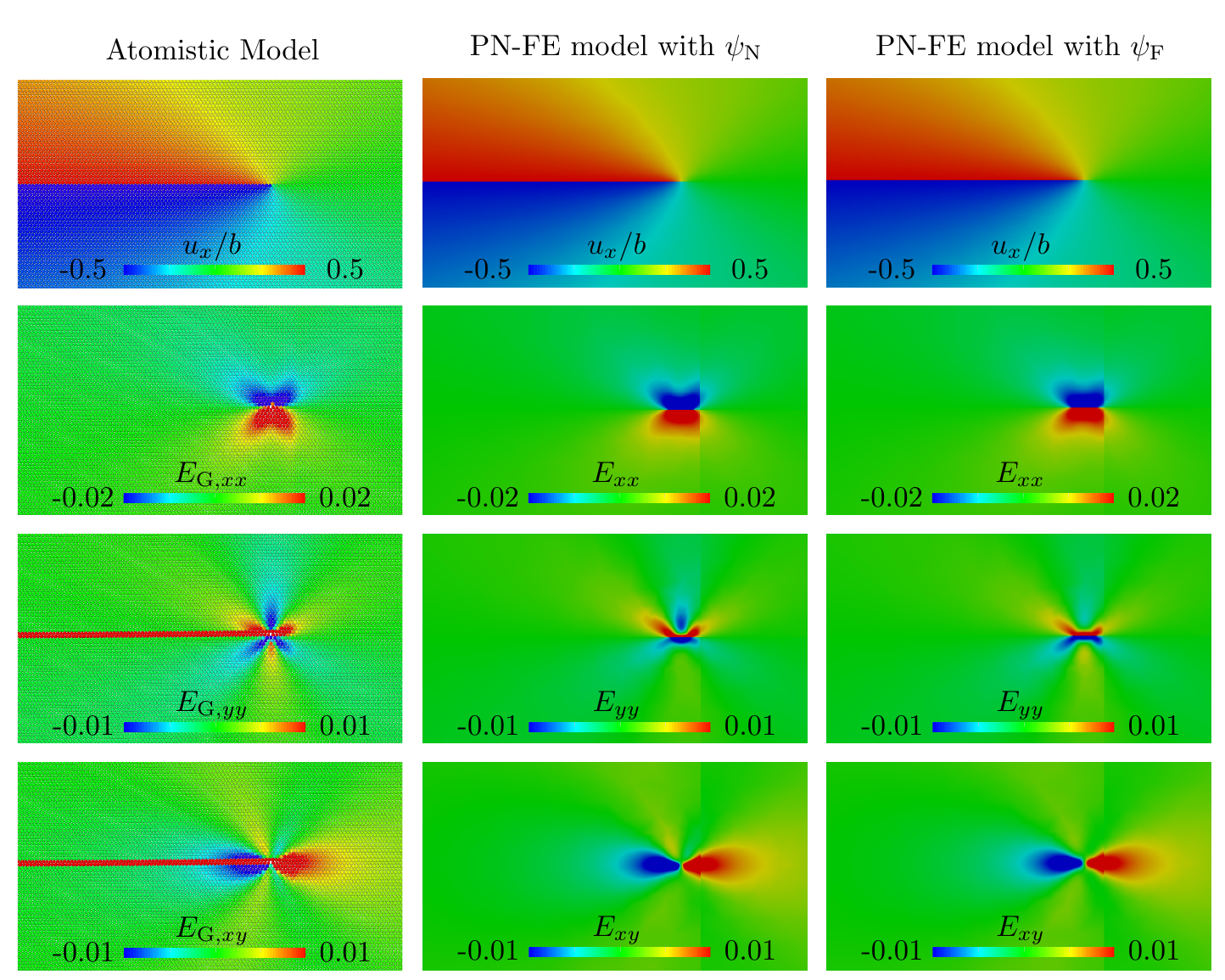}	
	\caption{Displacement ($x$-component) and strain field of the atomistic model compared to the PN-FE model with the Fourier glide plane potential~$\psi_{\mathrm{F}}$ and the coupled numerical glide plane potential~$\psi_{\mathrm{N}}$ at~$\tau=0.00473\,\mu^{\mathrm{A}}$.}
	\label{fig:field_plot-step-71}
\end{figure}

The local displacement and strain fields are next evaluated for the MS model and the PN-FE model with selected glide plane models. Considered are the coupled numerical potential ($\psi_{\mathrm{N}}$) and the Fourier potential ($\psi_{\mathrm{F}}$), which shows relatively good agreement in $\Delta_t$ and $T_t$ but is constrained in $\Delta_n$. For the atomistic model, the displacements are plotted in terms of the change of atom positions. The strains correspond to the Green--Lagrange strain tensor~$\boldsymbol{E}_{\mathrm{G}}$ obtained from the deformation gradient, which is calculated by a local least squares fit of the displacements of neighbouring atoms relative to the central one with respect to a homogeneous deformation gradient, as presented by~\cite{Shimizu2007}. The post-processing is performed in Ovito~\citep{Stukowski2010}. Because the strain fields are averaged quantities in MS, unreliable values are obtained within the dislocation core region. The strains are, therefore, to be evaluated with caution within the core regions. Accordingly, the used colour scales are cropped. For the PN-FE model, either directly the nodal quantities (for displacements) or the element averaged nodal quantities (for infinitesimal strains~$\bs{E}$) are shown. Due to~(i) the questionable accuracy of strains within the core region in the atomistic model, and~(ii) the difference between the geometrical non-linearity  of the atomistic model versus linearity of the PN-FE model, the obtained results can only be compared qualitatively. All figures span the window~$180\,a_0\le x\le 290\,a_0, -30\, a_0\le y \le 30\, a_0$. The resulting quantities are shown in Figure~\ref{fig:field_plot-step-71}.\par
The comparison shows a good agreement between the PN-FE model ($\psi_{\mathrm{N}}$ and $\psi_{\mathrm{F}}$) and the atomistic model. Notwithstanding its simplicity, the PN-FE model is able to capture the quite complex displacement and strain fields rather well. In both models, MS and PN-FE, similar discontinuities at the phase boundary are noticeable, e.g., in~$E_{xx}$, $E_{\mathrm{G},xx}$, and~$E_{xy}$, $E_{\mathrm{G},xy}$. These discontinuities are due to the change in material properties across the phase boundary, and hence are more pronounced in the PN-FE model, in which this boundary is sharper. Note that the large strains along the glide plane in the MS model ($E_{\mathrm{G},yy}$ and $E_{\mathrm{G},xy}$) stem from the large derivative of the displacement in the~$y$ direction, thus leading to high concentrated strain components across the glide plane. The difference between the MS and PN-FE model comes from the fact that in MS the total strain is shown, whereas in the PN-FE only the elastic strain is plotted. As a consequence, in the PN-FE model relative displacements across the glide plane are absorbed in the disregistry and thus do not contribute to the (bulk) strain.\par
Despite the relatively good agreement, some differences between the atomistic and the PN-FE model with~$\psi_{\mathrm{N}}$ are present, especially for~$E_{yy}$ and $E_{\mathrm{G},yy}$ below the glide plane. The Fourier glide plane potential~$\psi_{\mathrm{F}}$, on the contrary, achieves better qualitative agreement with the atomistic model. This, however, occurs in relation with the glide plane constraint~$\Delta_n = 0$, which limits the dislocation relaxation out of plane, and hence may not be misinterpreted as a higher accuracy of~$\psi_{\mathrm{F}}$. Both glide plane potentials show minor deviations for~$E_{xx}$ and~$E_{\mathrm{G},xx}$.\par
\subsection{Dislocation pile-up}
\label{sect:disl-pu}
\subsubsection{Disregistry profile}
\begin{figure}[htbp]
	\centering
	\begin{subfigure}{0.49\textwidth}
		\centering
		\caption{}
		\includegraphics[width=1.\textwidth]{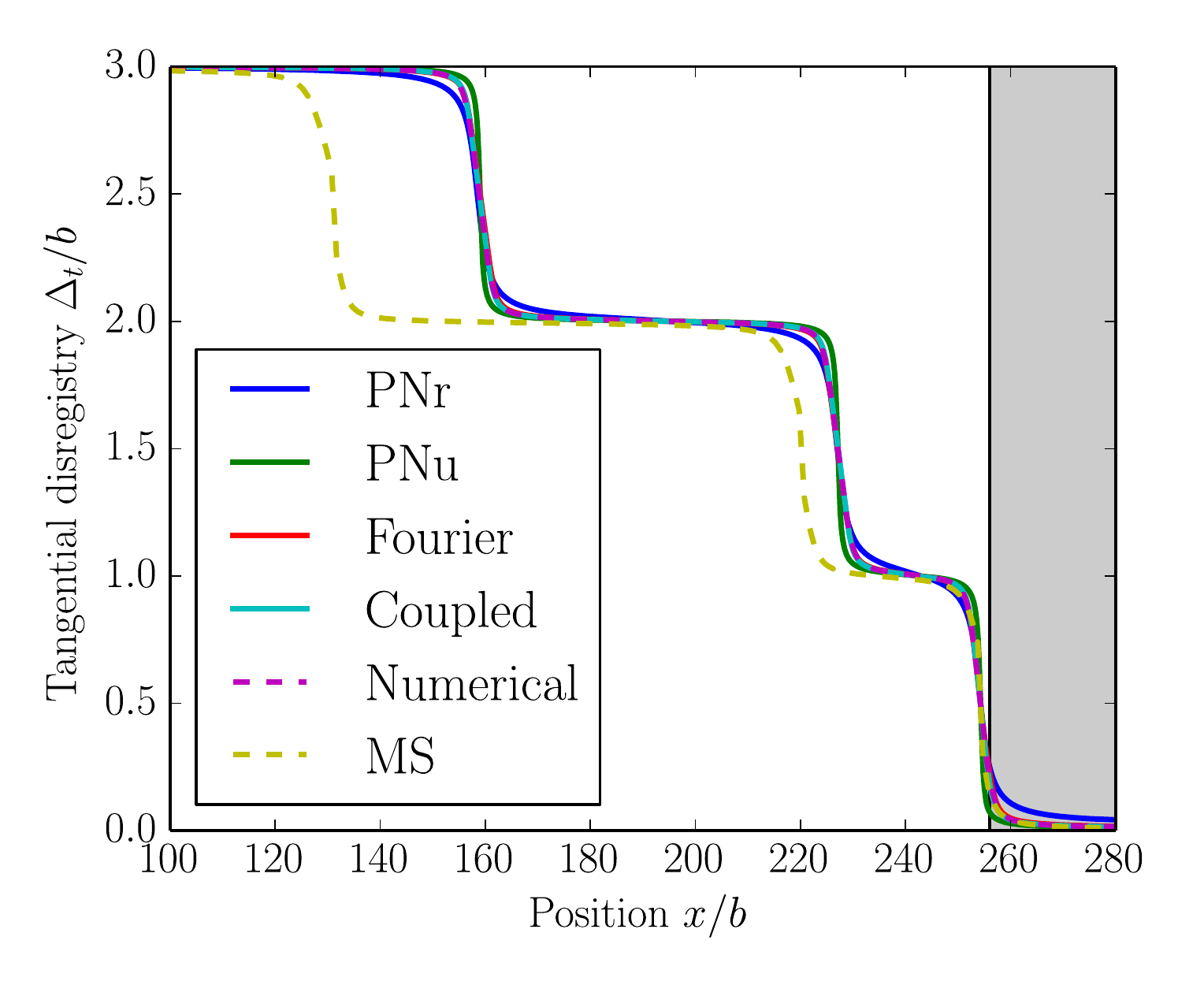}
		\label{fig:Disregistry-tan-step-159}
	\end{subfigure}	
	\begin{subfigure}{0.49\textwidth}
		\centering
		\caption{}
		\includegraphics[width=1.\textwidth]{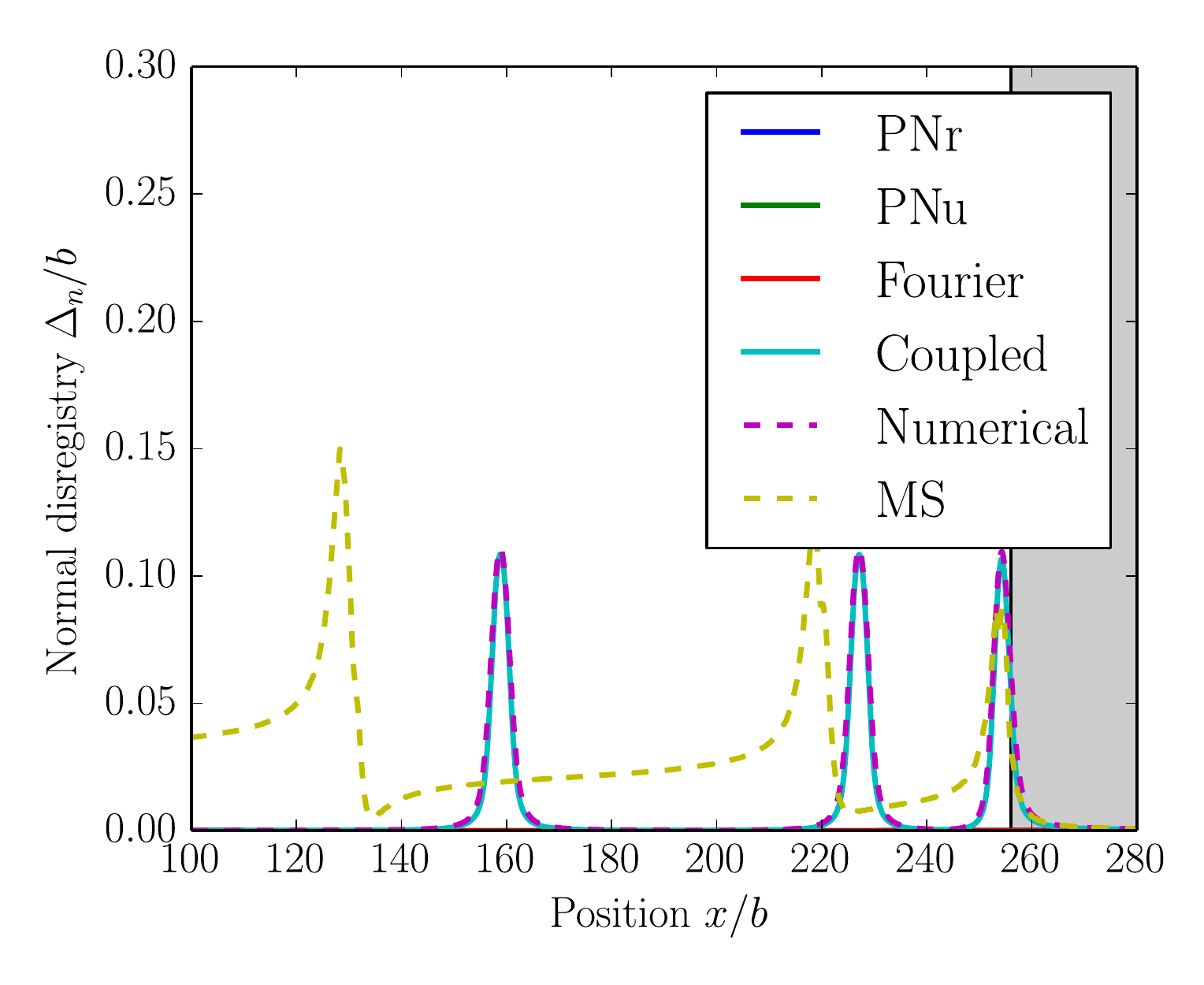}
		\label{fig:Disregistry-norm-step-159}
	\end{subfigure}
	\caption{Disregistry profile of the different glide plane potentials in comparison with MS at $\tau=0.0106\,\mu^{\mathrm{A}}$: (a) tangential disregistry~$\Delta_t$; (b) normal disregistry~$\Delta_n$.}
	\label{fig:Disregistry-step-159}
\end{figure}

To study the model responses under a pile-up configuration, an externally applied shear load of~$\tau=0.0106\,\mu^{\mathrm{A}}$ is considered, at which in the MS model and the PN-FE model a three-dislocation pile-up has formed. The disregistry profiles for the pile-up configuration are plotted in Figure \ref{fig:Disregistry-step-159}. \par
A mismatch of the dislocation positions between the MS model and the PN-FE model is apparent. In consideration of the equal long-range stresses (recall Figure \ref{fig:Shear-Traction-step-71}), this difference can be related to an energy barrier that needs to be overcome to displace the dislocation by one atomic spacing -- the so-called Peierls barrier -- originating from the lattice discreteness. This additional resistance against dislocation motion results in a larger pile-up length than in the continuum PN-FE model, where no Peierls barrier is present, recall Section~\ref{sect:pnfe} and the discussion on discretization of the PN-FE model therein. Although the magnitude of the Peierls barrier has not been computed explicitly for the atomistic system, the conjecture on its effect has been verified by a set of additional numerical simulations (not reported here for conciseness), in which the individual dislocations were initialized at different spatial positions on the glide plane~$\Gamma_{\mathrm{gp}}$. The observed scatter in the resulting equilibrium dislocation positions, $x_j$, $j = 1,2,3$, corresponds in magnitude to the differences observed between the MS and PN-FE model in Fig.~\ref{fig:Disregistry-step-159}, thus ascertaining that the Peierls barrier is at the root of the difference between the two methods. Again, an artificial contribution is included in the calculation of the normal disregistry $\Delta_n$ which naturally increases with a larger $\Delta_t$ (and a larger number of dislocations).\par
The comparison of the different potentials shows a negligible influence of the adopted glide plane potential on the dislocation positions, which is in alignment with the equality of the long-range repulsive shear stresses for sufficiently separated dislocations ($>5b$). An evaluation of the glide plane shear tractions, as done for the single dislocation case (recall Figure \ref{fig:Shear-Traction-step-71}), is omitted here as it does not provide any additional insight.\par
\subsubsection{Displacement and strain field}
\begin{figure}[htbp]
	\centering
	\includegraphics[scale=1]{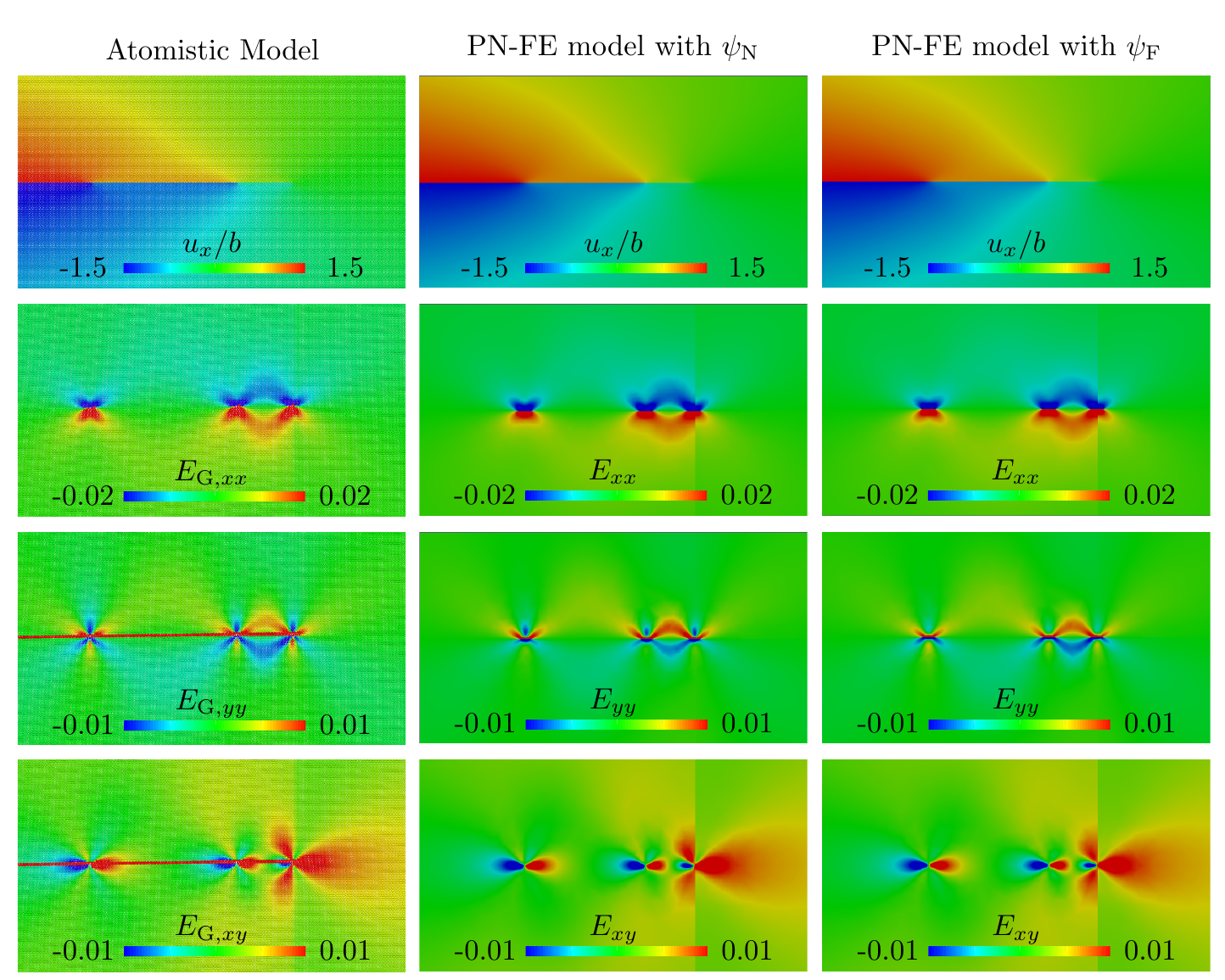}	
	\caption{Displacement ($x$-component) and strain field of the atomistic model compared to the PN-FE model with the Fourier glide plane potential~$\psi_{\mathrm{F}}$ and the coupled numerical glide plane potential~$\psi_{\mathrm{N}}$ at~$\tau=0.0106\,\mu^{\mathrm{A}}$. Note the different colour scale for $u_x$ compared to Figure~\ref{fig:field_plot-step-71}, reflecting the presence of three dislocations (versus one).}
	\label{fig:field_plot-step-159}
\end{figure}

The corresponding displacement and strain fields in~$100\,a_0\le x\le 320\,a_0, -60\, a_0\le y \le 60\, a_0$ are plotted in Figure \ref{fig:field_plot-step-159} for the MS model and the PN-FE model with the coupled numerical potential ($\psi_{\mathrm{N}}$) and the Fourier potential ($\psi_{\mathrm{F}}$). In addition to the insight obtained for the single dislocation case (recall Section \ref{sect:single_disl_fields}), the additional discrepancy in dislocation positions (between the atomistic and the PN-FE models) is apparent here from the slight difference in the displacement and strain fields. The general agreement between the different fields, however, is quite satisfactory.
\subsection{Pile-up evolution and transmission}
\label{sect:dislocations}
In this section, the pile-up formation is studied for the MS and the PN-FE model in detail, i.e., from the initially dislocation-free lattice up to the transmission of the leading dislocation. The PN-FE model with $\psi_{\mathrm{N}}$ is first compared with the atomistic model; a discussion on the influence of the adopted glide plane potential follows thereafter. \par
\begin{figure}[htbp]
	\centering
	\begin{subfigure}{0.49\textwidth}
		\centering
		\caption{}
		\includegraphics[width=1.\textwidth]{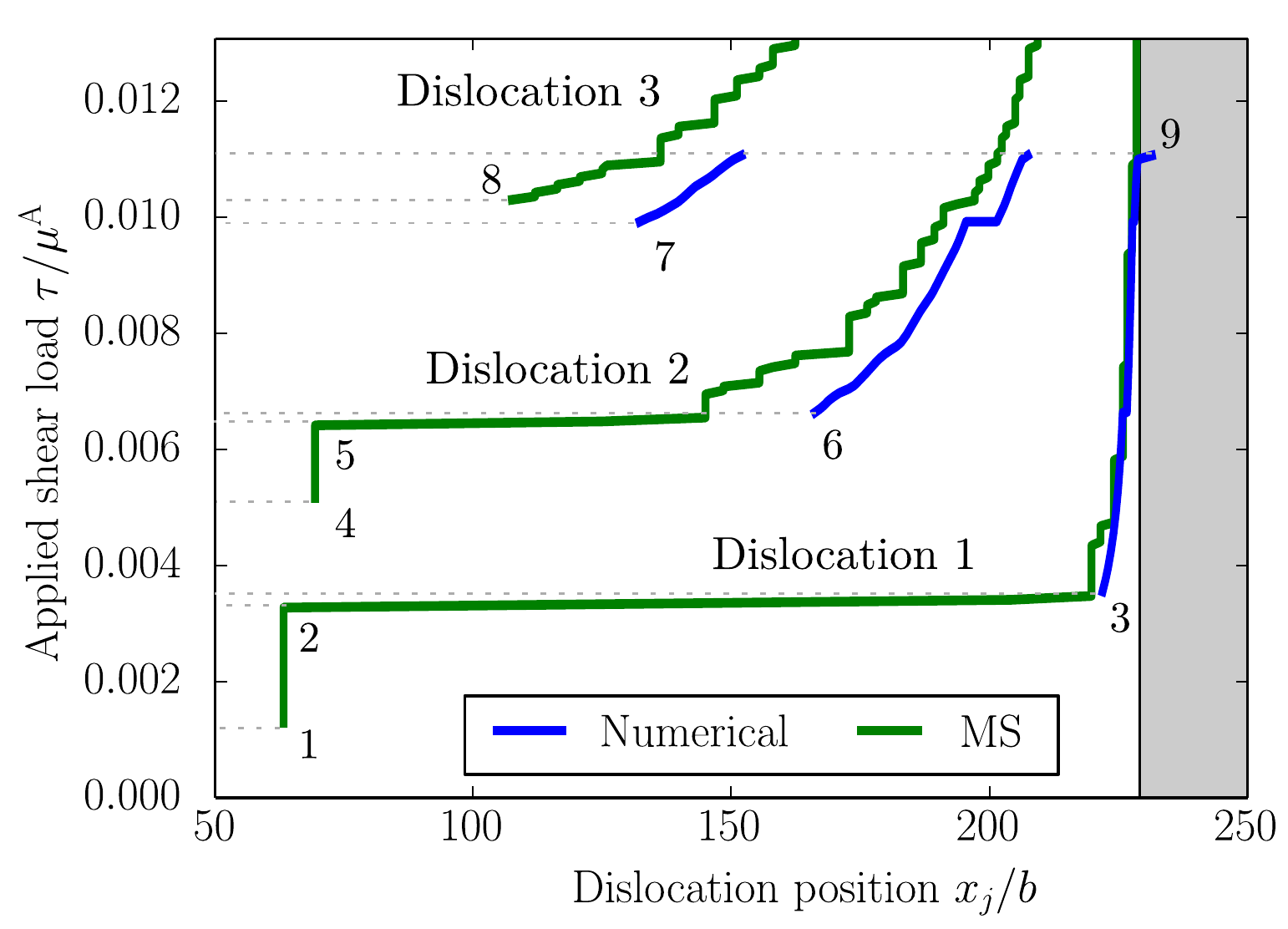}
		\label{fig:global_response_pos}
	\end{subfigure}	
	\begin{subfigure}{0.49\textwidth}
		\centering
		\caption{}
		\includegraphics[width=1.\textwidth]{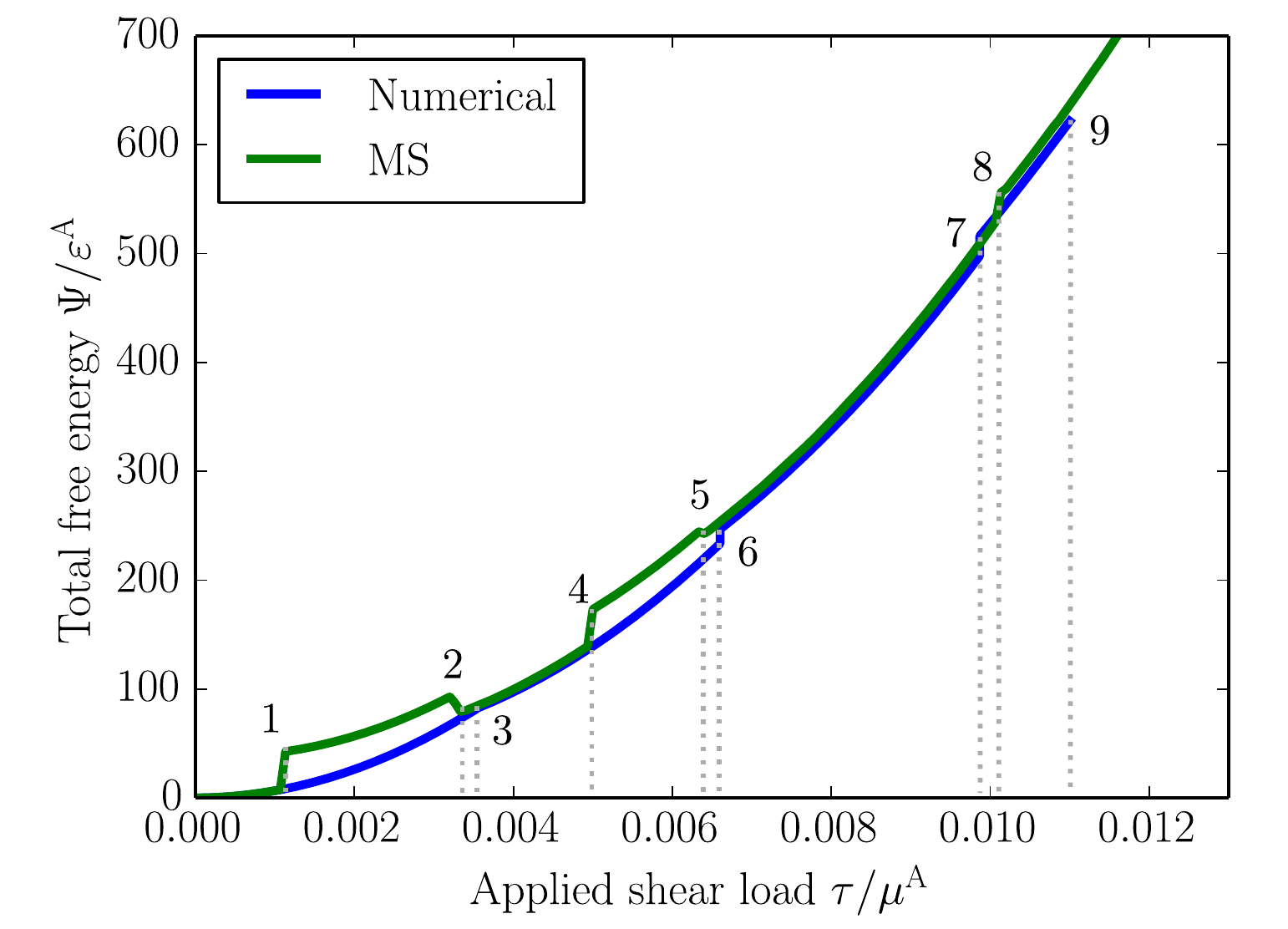}
		\label{fig:global_response_energy}
	\end{subfigure}
	\caption{Global model response to the formation and evolution of a three-dislocation pile-up under increasing externally applied shear load~$\tau$ for the atomistic model (MS) and the PN-FE model with the coupled numerical glide plane potential (Numerical): (a) evolution of the dislocation positions~$x_j$ and (b) evolution of the energy. Transmission occurs in the PN-FE model at an externally applied shear load of~$\tau = 0.011\,\mu^{\mathrm{A}}$. The grey area indicates Phase~B.}
	\label{fig:global_response}
\end{figure}

By subjecting the considered microstructure to a monotonically increasing external shear deformation (corresponding to the shear load $\tau$), stable dislocations are successively nucleated in both models. As a consequence of that applied shear load, the nucleated dislocations tend to move towards the phase boundary where, as a result of the present phase contrast, they pile up. In the simulations, the pseudo time~$t$ is updated by increments of~$1/450$ to reach the target strain~$\overline{\tau}=0.03\,\mu^{\mathrm{A}}$, until dislocation transmission is recorded. The detailed evolution of the dislocation positions as a function of the externally applied shear load~$\tau$ is shown for the MS and the PN-FE model ($\psi_{\mathrm{N}}$) in Figure \ref{fig:global_response_pos}, where $x_j$ represents the horizontal coordinate of dislocation $j$. The corresponding evolution of the total free energy $\Psi$ is illustrated in Figure \ref{fig:global_response_energy}, with~$\Psi=\mathcal{V}(\utilde{r})-\mathcal{V}(\utilde{r}_0)$ for the MS model. The specific model evolutions, in terms of dislocation positions and total free energy can be understood as follows.\par
Initially, both models are defect free and behave (nearly) linear elastic. Only after the external shear deformation has increased sufficiently, dislocation nucleation is triggered, at~$\tau=0.0012\,\mu^{\mathrm{A}}$ for the atomistic system; in the PN-FE model, the criterion for dislocation nucleation is not fulfilled yet. By nucleating the dislocation, a significant amount of additional energy is introduced into the system (cf. Figure~\ref{fig:global_response_energy}). At the point of nucleation (i.e., Point~$1$ in Figure~\ref{fig:global_response}), the stable and unstable equilibrium positions of the dislocation lie close to each other, meaning also that multiple local energy minima exist. The individual local minima are separated by the Peierls barriers. This occurrence of multiple local minima, and thus Peierls barrier, does not exist in the continuum formulation of the PN-FE model. Hence, no stable dislocations are nucleated in the PN-FE model yet. It is also to be noted that in a Molecular Dynamics (MD) simulation the dislocation may overcome the barrier due to thermal activation and annihilate (recall the dipole configuration). Nevertheless, due to the presence of that barrier in the MS model the externally applied shear load~$\tau$ needs to be increased to propagate the dislocation further towards the phase boundary, into its new equilibrium position ($\tau=0.0033\,\mu^{\mathrm{A}}$ -- Point 2). Subsequently, a sudden motion occurs, which is reflected in the energy curve by a large drop in the energy to a slightly higher level compared to the PN-FE model. Shortly thereafter ($\tau = 0.0035\,\mu^{\mathrm{A}}$ -- Point~$3$), the shear deformation suffices to trigger dislocation nucleation in the PN-FE model close to the phase boundary. Here, the energy introduced to nucleate the dislocation equals the amount of energy by which the system relaxes due to the presence of the dipole. Not a jump, but rather a kink is therefore observed in~$\Psi(\tau)$. At this point, a slight difference in $\Psi$ between the MS and the PN-FE model reveals a minor underestimation of the dislocation-induced energy by the PN-FE model. Upon further increasing the externally applied shear load, both models exhibit similar behaviour in terms of the energy and position of the first dislocation. The minor difference in dislocation position is explained by the presence of the Peierls barrier, which leads to a step-wise motion of the dislocation in the atomistic model with steps of approximately one Burgers vector~$b$ -- a phenomenon that is not present in the PN-FE model due to its continuous nature.\par
A second dislocation is nucleated in the atomistic model at~$\tau=0.0051\,\mu^{\mathrm{A}}$ (Point~$4$). An energy increase is again required to nucleate it, and the dislocation remains temporarily in its nucleation position until the applied shear load reaches the value of~$\tau=0.0065\,\mu^{\mathrm{A}}$. Again, in MD simulations this nucleated dislocation may not be stable. By further increasing the applied shear load, the dislocation moves towards the phase boundary where it piles up, leading to a relaxation of the system and a small drop in energy (Point~$5$). At approximately~$\tau=0.0065\,\mu^{\mathrm{A}}$ (Point~$6$), the second dislocation is nucleated in the PN-FE model, which now requires some energy, resulting in a small jump. With the second dislocation the difference in energy between both models increases, as opposed to where only one dislocation was present, due to the marginal underestimation of the dislocation-induced energy in the PN-FE model. The comparison of the dislocation positions (Figure \ref{fig:global_response_pos}) furthermore reveals a small deviation between both models which, similar to the step-wise dislocation motion, can be attributed to the Peierls barrier. \par
With the nucleation of the third dislocation the mechanism changes. Not only does the dislocation nucleate in the atomistic model at a higher shear load ($\tau=0.0103\,\mu^{\mathrm{A}}$ -- Point~$8$) compared to the PN-FE model ($\tau=0.0098\,\mu^{\mathrm{A}}$ -- Point~$7$), it is also immediately mobile. Both the later nucleation and mobility result from the already existing dislocation structure.\par 
In the atomistic model, the dislocation pile-up is at all stages less compressed and it hence induces a higher long-range backstress compared to the PN-FE model, triggering a later nucleation of subsequent dislocations. With the nucleation of the third dislocation a jump in the position of the second dislocation and the absence of the motion barrier for the third dislocation (in the MS model) is apparent, which can be explained as follows. At the beginning of the nucleation step the third dislocation is introduced at position $x_3$ where it initially exerts a repulsive shear stress on the two-dislocation pile-up. Hence, in the course of minimising the total free energy (at this step), the pile-up is compressed, which in turn lowers the repulsive shear stress of the pile-up on the nucleated dislocation. As a result, the third dislocation experiences a driving force towards the phase boundary where, at the end of the nucleation step, it settles down in its new position $x_3$. Despite the present differences between both models, the general trend of the MS model, in terms of dislocation positions and total free energy, is followed qualitatively well and to some degree even quantitatively.\par
Eventually, the leading dislocation is transmitted in the PN-FE model at an externally applied shear load of $\tau=0.011\,\mu^{\mathrm{A}}$. In the atomistic model, a significantly higher stress level of $\tau=0.0167\,\mu^{\mathrm{A}}$ is required to reach transmission, at which the size of the pile-up has already increased by one dislocation. For the sake of conciseness and to facilitate a sharp comparison between both models this is not included in Figure~\ref{fig:global_response}. The cause of the different stress required for dislocation transmission lies (i) within the strong influence of the Peierls barrier on the pile-up length, and (ii) within the complex interaction of the dislocation core with the phase boundary during its transmission across the phase boundary. Due to the strong approximations made on the core behaviour in the PN-FE model (i.e., continuum description, small deformations, linear elasticity, local glide plane potential), it is not able to capture the process of dislocation transmission accurately.\par
\begin{figure}
	\centering
	\includegraphics[width = 0.5\linewidth]{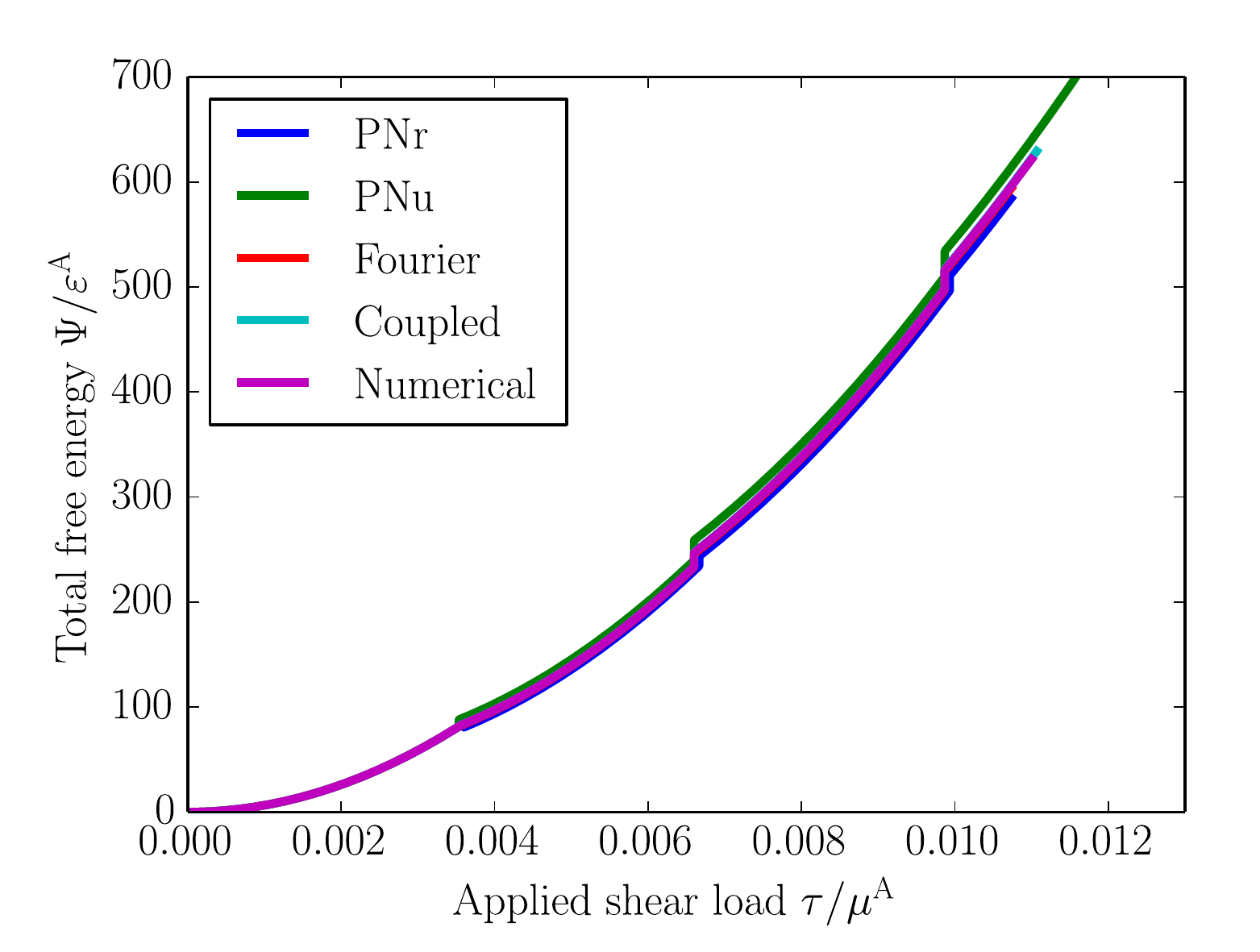}
	\caption{Comparison of the global model response in terms of the energy under increasing externally applied shear load~$\tau$ for the different glide plane potentials of the PN-FE model. }
	\label{fig:Energy_plot_comp}
\end{figure}

A comparison of the energy evolution obtained with the different glide plane potentials, plotted in Figure~\ref{fig:Energy_plot_comp}, shows a small difference up to a shear load of~$\tau\approx 0.011\,\mu^{\mathrm{A}}$. Similarly to the dislocation glide plane shear tractions (recall Figure \ref{fig:Shear-Traction-step-71}), PNr (with $\gamma_{\mathrm{us}}^{(\mathrm{r})}$) relates to the lowest total free energy and PNu (with $\gamma_{\mathrm{us}}^{(\mathrm{u})}$) to the highest, whereas the energy for the Fourier, the coupled analytical and coupled numerical potentials exhibit a (nearly) matching total free energy. Most glide plane potentials invoke dislocation transmission at~$\tau=(0.0109\pm 0.00017)\,\mu^{\mathrm{A}}$. Only with $\psi_{\mathrm{PN},\mathrm{u}}$ (fitted to~$\gamma_{\mathrm{us}}^{(\mathrm{u})}$) dislocation transmission in triggered at a strongly increased shear load with~$\tau=0.0214\,\mu^{\mathrm{A}}$, where already six dislocations have been nucleated. This illustrates the significant influence of~$\gamma_{\mathrm{us}}$ on dislocation obstruction at the phase boundary.\par

%
%
\section{Summary and discussion}
\label{section:Discussion}
This paper presented a qualitative and quantitative comparison of the PN-FE model against equivalent atomistic simulations in terms of edge dislocation behaviour using a 2D two-phase microstructure. Different glide plane potentials were considered to study their influence on the dislocation structure (i.e., disregistry profile) and on dislocation transmission across the phase boundary. Although the continuous PN-FE methodology is based on numerous simplifications, it is able to capture the atomistic response to a good extent in terms of dislocation positions, disregistry profile and strain field. This exemplifies that the PN-FE model is able to capture the long-range influence of a second phase on the dislocation behaviour fairly well. Some quantitative differences between the atomistic and PN-FE model are present, originating from a distinct dislocation core behaviour in terms of the disregistry profile, the pile-up length ($\approx 15\%$) and the shear load which is required for dislocation transmission. These differences can be related to the simplicity of the PN-FE model as follows: (i)~linear elasticity describes the behaviour of the bulk above and below the glide plane; (ii)~the problem is solved in an infinitesimal strain framework; (iii)~the material properties are homogeneous within the individual phases, with a jump across the interface; (iv)~the behaviour of a discrete (atomic) system is adopted in a continuum formulation, which thus cannot capture the discrete nature of the Peierls barrier; and (v)~the effect of non-locality is neglected. Although it is possible to advance the PN-FE model to higher complexity, its simplicity and computational efficiency would be compromised.\par
The comparison of the different glide plane potentials of the PN-FE model showed that the amplitude of the glide plane potential has a significant influence on dislocation transmission. Glide plane potentials fitted to the relaxed unstable stacking fault energy~$\gamma_{\mathrm{us}}^{(\mathrm{r})}$ invoked dislocation transmission at a similar externally applied shear deformation compared to atomistic simulation. The glide plane model fitted to the unrelaxed unstable stacking fault energy~$\gamma_{\mathrm{us}}^{(\mathrm{u})}$, on the contrary, required a much higher shear deformation to trigger dislocation transmission. In terms of tangential disregistry profile $\Delta_t$, the highest accuracy (compared to the reference numerical potential~$\psi_{\mathrm{N}}(\Delta_n,\Delta_t)$) was obtained by the coupled potential~$\psi_{\mathrm{C}}(\Delta_n,\Delta_t)$ and the Fourier series based potential~$\psi_{\mathrm{F}}(\Delta_t)$, although the latter disregards the normal disregistry $\Delta_n$. The normal disregistry~$\Delta_n$ is only incorporated in the coupled model~$\psi_{\mathrm{C}}$ and matches adequately with the reference model~$\psi_{\mathrm{N}}$.\par
Although numerous simplifications are adopted in the PN-FE model, it has proven to be an elegant and computationally efficient approach to study edge dislocation interactions with phase boundaries. It enables to study isolated mechanisms of the underlying physics, since only the key mechanisms employed in the model are activated. 
%
%
\section*{Acknowledgements}
We would like to thank Prof. Bob Svendsen from RWTH Aachen University and the Max Planck Institute for Iron Research for useful discussions on the present work. This research has been supported by the Tata Steel Europe through the Materials innovation institute (M2i), project no.~S22.2.1349a, and by the Netherlands Organisation for Scientific Research (NWO), grant no.~STW 13358. Financial support of this research received from the Czech Science Foundation (GA\v{C}R) under project no.~17-04150J is also gratefully acknowledged.
%
%

\bibliography{Paper3_formated}

\end{document}